\documentstyle [preprint,prb,eqsecnum,aps]{revtex}
\begin {document}
\draft
\title {Frustrated Bonds and Long Range Order in Quasi-2D
Magnets.}
\author{I.~Ya.~Korenblit }
\address {School of Physics and Astronomy, Raymond and Beverly
Sackler Faculty of Exact Sciences,
\hfil\break Tel Aviv University, Tel Aviv 69978, Israel}
\maketitle
\begin {abstract}
We employ the  Schwinger boson mean-field approach to
study the effects of arbitrary frustrated
 bonds and plaquettes (formed from four frustrated bonds)
in two-dimensional
ferro- and antiferromagnets on the spin-wave spectrum and the correlation
length
 at finite temperatures. We distinguish between strongly
frustrated bonds (plaquettes), when the frustrated coupling $J^\prime$
exceeds the spin canting threshold $J_c$, and weakly frustrated
bonds (plaquettes), with $J^\prime <J_c, (J_c-J^\prime)/J_c\sim 1$.
 It is shown that in antiferromagnets
the amplitude of
spin-wave scattering on strongly frustrated bonds or plaquettes
 grows with the decrease of the
temperature.
A small amount of such defects
reduces significantly the spin-wave stiffness and
the correlation length at low temperatures. As a result, the
quasi-2D N\'eel temperature is sharply suppressed.
Quantum fluctuations are also considered and their
effect on the spin-wave spectrum is shown to be of the
order of $(2S)^{-2}\ln^{-1}2S$
in the large spin limit.
For weakly frustrated (nonfrustrated) defect bonds (plaquettes)
 the spin-wave stiffness renormalization is of the
order of the dopant concentration and does not depend on the
temperature.
 The results account for the observed properties of
 doped
 quasi-2D
$La_2CuO_{4+x}$.  The frustrated bonds in 2D ferromagnets act
in the same way as in antiferromagnets, while the effect of frustrated
plaquettes is weak and temperature independent.

\end{abstract}

\section{Introduction}
\label{sec:intro}
The discovery of high $T_c$ superconductivity in the doped cuprates
$La_2 CuO_{4+x}$ and $YBa_2 Cu_3 O_{6+x}$ enhanced the interest
in the study of two-dimensional (2D) quantum antiferromagnets. It is now well
established
that the system of localized $Cu^{2+}$ spins in both families (near $x$=
0) is described
by the 2D Heisenberg Hamiltonian, and a small coupling between $Cu^{2+}$ layers
leads to the formation of a three-dimensional N\'eel ordered state at
temperatures much smaller than the intraplane exchange interaction
energy. \cite{bir,keim1} Above  the N\'eel temperature $T_N$ one observes
2D antiferromagnetic correlations. To leading order, the correlation
length $\xi$ in undoped samples follows \cite{keim1,grev} the theoretically
predicted \cite{pol,chak,ar} law
\begin{equation}
 \xi \propto \exp(A/T).\label{cor}
\end{equation}
The calculation by Chakravarty {\it et al.},\cite{chak} based on
the quantum nonlinear
sigma model ($QNL\sigma M$), showed that $A = 2\pi \rho_s $, where
$\rho_s$ is the spin stiffness constant. The preexponentional factor
in the two-loop approximation does not depend on the temperature
up to terms of the order of $T/2\pi \rho_s$  in antiferromagnets and
scales as $T^{1/2}$ in ferromagnets. A very good quantitative agreement between
the $QNL\sigma M$ correlation length and the results of a neutron scattering
study of the model $2D$, $S=1/2$, square-lattice Heisenberg antiferromagnet
$Sr_2CuO_2Cl_2$ was recently obtained in a surprisingly wide temperature range
$T_N<T<2\pi\rho_s$.\cite{grev}

Another approach to low-dimensional ($D =1,2$) magnets has been proposed
by Arovas and Auerbach.\cite{ar}  They employed the Schwinger boson
representation  for spins in a $SU(N)$ generalization of the Heisenberg
model. Their expression for the correlation length corresponds to the
one-loop aproximation in the $QNL\sigma M$.
 The constant $A$ in Eq.~(\ref{cor})  is $2\pi JS^2$ in the limit of
large $S$, and $J$ stands for the intraplane coupling constant.

The doping has a drastic effect on the magnetic properties of the
cuprates. Even a very small dopant concentration, which indroduces
holes residing on the oxygen orbitals in the $CuO_2$ planes,
substantially reduces
the N\'eel temperature.\cite{bir,keim1,cho1,cho2}
 In $La_2 CuO_4$, doped with strontium or with
 excess
oxygen, the long range order disappears at doping concentration as
small as $2\%$.
 It has been shown \cite{keim1,endo,yam}
that in samples of $La_2 CuO_{4+x}$ with a small concentration of defects
(of the order or less than 1\%) the correlation length varies little with
doping at
very high temperatures 500 - 600 K, however the growth of the spin
correlations is increasingly inhibited by the defects as the temperature
is decreased.

In this paper we shall focus on the properties of doped 2D magnets.
Doping with excess oxygen or with strontium creates holes on the
oxygen sites in the $CuO_2$ planes.\cite{em} Since the
Cu-O distance is half the Cu-Cu distance,
the exchange interaction, $J_\sigma$, of the hole spin with the
two nearest-neighbor
Cu spins is much larger than the Cu-Cu exchange, $J$.
For either sign of $J_{\sigma}$  the hole spin would thus like to be
parallel (antiparallel) to the two neighboring Cu spins. Therefore
the oxygen hole
introduces an effective {\it ferromagnetic} coupling,
$J^\prime$, between its two
Cu nighbors, with $J^\prime =O(\mid J_{\sigma}\mid)\gg J$.
This was the basic idea of the frustration model of Aharony {\it et al.}.
\cite{ahar}

 Motivated by these arguments
 and the results of the
experimental investigation of $La_2 CuO_{4+x}$ and $La_{2-x}Sr_xCuO_4$,
 \cite{bir,keim1,endo,yam,keim2}
 we shall consider two
 simple models of defects, namely, frustrated bonds and
frustrated plaquettes.
We hope
that the main qualitative results, presented in this paper, survive in
the case of more complicated frustrating defects.

The effect of randomly distributed frustrated bonds on the Heisenberg ferro-
and antiferromagnets in the ordered state has
been investigated by many authors.\cite{gins,feig,vann,am,ls} It has been shown
that
there is a threshold energy of the  frustrated bond,
\begin{equation}
J_c = (z/2 -1)J,\label{thresh}
 \end{equation}
which determines the local stability of the system. Here
 $z$ is the number of the nearest neighbors. For a 2D
 square lattice, $z=4$ and $J_c = J$.
 If the (positive) energy
of the frustrated bond $J^\prime$ exceeds $ J_c$,
 the two spins
connected by the frustrated bond cant. In the classical
limit the canted spins act on the magnetic background as a
dipole.\cite{ahar,vill} As a result, the spins around the ferromagnetic bond
are also canted, and the canting angle in a 2D Heisenberg magnet
decays at large distances $r$ from the defect as $r^{-1}$.

The quantum fluctuations of the spins, connected by the frustrated
bond, have been calculated for $J^\prime < J_c$ in the framework
of the linear spin wave theory. \cite{am,ls} These fluctuations diverge when
$J^\prime \to J_c^- $ , which reflects the breakdown of the linear spin
wave approach.

The effect of localized holes on the properties of quasi-2D
antiferromagnets at finite temperatures has been studied by Glazman
and Ioselevich.\cite{gi} A classical approach, based on the dipole model
of Aharony {\it et al.},\cite{ahar} led them to the
conclusion that the renormalized stiffness is a function of
$(x/T)$, where $x$ is the concentration of localized holes.

The model of a single frustrated bond is reasonable for samples
doped with excess oxygen. In contrast, $Sr$ doping replaces a trivalent
$La^{3+}$ ion with a divalent $Sr^{2+}$ ion in the plane above
a $CuO_2$ plane. The Coulomb potential that pins the hole in
the $CuO_2$ plane is then due to the $Sr^{2+}$ ions, which project onto
the centers of the copper-oxygen plaquettes. Therefore the holes are
localized on small regions around the centers of the above plaquettes.
The simplest defect model of this kind is a plaquette formed by
four frustrated bonds.\cite{pu,good} The canted spins at the corners
of the plaquette act on the long-range order parameter (at
$T\rightarrow 0$) as a quadrupole rather than a dipole.
The energy spectrum of  the hole, localized on a plaquette, was
calculated in Refs.~\onlinecite{pu} and \onlinecite{good}.
 To our best knowledge
neither the critical value $J_c$ of the frustrated coupling (at T=0),
 nor the effect of the frustrated plaquette at finite temperatures,
has been considered till now.

In this paper we study the effects of defect (frustrated
and nonfrustrated) bonds and plaquettes
in 2D
ferro- and antiferromagnets on the spin wave spectrum and on the
correlation length at finite temperatures.
The Schwinger boson  mean-field approach (SBMFA)\cite{ar}
will be used. This method has been applied successfully to study
a variety of properties of low dimensional antiferromagnets.
\cite{keim2,chub,kop1,mill,kop2,ng,azz,zh,mim,matt}

 We show that there is a striking distinction
 between the effect
of strongly and weakly  frustrated defects on the behavior
of doped antiferromagnets.
 For strongly frustrated defects, with $t= (J^\prime-
J_c)/J_c>0$  ($J_c = J$ for
 frustrated bonds, $J_c = 0.376J$ for frustrated plaquettes),
 the renormalization
of the spin wave velocity $c$ and the correlation length in the
large spin limit (to leading linear order in $x$) is given by
\begin{equation}
{c(x)-c(0)\over c(0)} = -\alpha{4JS^2t\over T(1+t)}x, \end{equation}
\begin{equation}
\ln{\xi(x,T)\over\xi(x,0)}= -\alpha{8\pi J^2S^4t\over T^2(1+t)}x,
 \end{equation}
where the multiplier $\alpha$ depends on the type of the defect
($\alpha \approx 1$ for frustrated bonds, $\alpha\approx 3.5$
for frustrated plaquettes), and
decreases slightly
with the decrease of the temperature.
Hence, a  small amount of frustrated bonds or plaquettes
reduces significantly the spin-wave velocity and
the correlation length at low temperatures
 $T\ll 4JS^2 $, if $t$ is of the order or larger than unity.
The decrease of the 2D
correlation length with doping causes the strong supression
of the quasi-2D
long range magnetic order. The N\'eel temperature changes with the
dopant concentration as \begin{equation}
{T_N(x)-T_N(0)\over T_N(0)} \approx -\alpha{4JS^2\over T_N(0)}x.
\end{equation}
Because of the large factor $4JS^2\alpha/ T_N(0)$, the N\'eel temperature
extrapolates to zero at a doping concentration which is much
smaller than the percolation threshold.

The quantum fluctuations, associated with
the frustrated bonds, are relatively large near the threshold
 $J^\prime = J_c$. However, when $J^\prime \gg J_c$, they are
surprisingly small, of the order of $(2S)^{-2}\ln^{-1}2S$.
For weakly frustrated (nonfrustrated) defect bonds or plaquettes,
 with $ (J_c- J^\prime)/J_c
\approx 1$, the spin-wave stiffness renormalization is of the
order of the dopant concentration and does not depend on the
temperature.

The frustrated bonds in 2D ferromagnets act
in the same way as in antiferromagnets. However the effect of frustrated
plaquettes in ferromagnets is entirely different.
The renormalization of the spin-wave stiffness and, hence, of
the correlation length is not enhanced, and it is temperature independent
at both $J^\prime < J_c$ and $J^\prime > J_c$.

The above results account for the observed properties of doped quasi-2D
antiferromagnets.

The paper is organized as follows. In Sec.~\ref{sec:fero} the
SBMFA
is used to study the problem of a defect bond
in a 2D Heisenberg ferromagnet. We derive the mean-field Hamiltonian
and find the temperature dependence of the spin-wave scattering
amplitude on the defect bond.
 Then we treat the ferromagnet with
a low concentration of bond defects, and obtain the renormalization
of the spin-wave stiffness and the correlation length at small
temperatures $T\ll2\pi \rho_s$.
 In Sec.~\ref{sec:afer} we extend this
method to treat
2D antiferromagnets with defect bonds, and study the effect of
such defects on $T_N$. The plaquette type defects are considered in
Sec.~\ref{sec:plaq}.
In Sec.~\ref{sec:dis} we summarize our results, and compare them with the
experimental findings. Some comments are made concerning the
temperature and defect concentration dependence of the spin
wave velocities in  quasi-2D antiferromagnets at $T \ll T_N$. The canting of
the spins
at large distances from the strongly frustrated ($J^\prime>J_c$)
bond is addressed in Appendix A. In Appendix B we calculate the $T$-matrix for
the scattering of the Holstein-Primakoff spin waves on frustrated plaquettes at
$T=0$ and $J^\prime < J_c$.

\section{ Frustrated bonds in a 2D ferromagnet}\label{sec:fero}
\subsection{ The Hamiltonian}

We consider a Heisenberg ferromagnet on a square lattice with
nearest neighbor (nn) interactions only.
 Thus, in the standard Heisenberg Hamiltonian
\begin{equation}
 H = - \sum_{<ij>}J_{ij}\vec S_i \cdot \vec S_j ,
\end{equation}
the sum is over nn pairs, $<ij>$, and the
exchange interactions $J_{ij}$ are equal to $J$ (for most bonds) or
$-J^\prime$ (for a small concentration of bonds).

 In the Schwinger representation, each spin is replaced by two bosons:
\begin{equation}
 S_i^+ = a_i^\dagger b_i , ~~S_i^- = a_i b_i^\dagger ,~~ S_i^z =
{1\over 2}(a_i^\dagger a_i - b_i^\dagger b_i),
\end{equation}
together with the constraints
\begin{equation}
{1\over2} (a_i^\dagger a_i +
b_i^\dagger b_i) = S. \label{const}
\end{equation}
 After this transformation is
performed, the Hamiltonian can be written as\cite{ar}
\begin{equation}
  H = -2\sum_{<ij>}J_{ij} :B_{ij}^\dagger B_{ij}:  +
\sum_i \lambda_i (a_i^\dagger a_i + b_i^\dagger b_i) ,\label{h}
\end {equation}
where $B_{ij}^\dagger = {1\over2} (a_i^\dagger a_j + b_i^\dagger b_j)$,
and :~: denotes normal ordering.
The second term in this expression is added to take into account the
constraint (\ref{const}),
 $\lambda_i$ are the Lagrange multipliers. An important
approximation in the mean field theory of Arovas and Auerbach \cite{ar} is
that the constraints (\ref{const}) are imposed on the {\it average}, i.e. they
are taken into account by introducing a single Lagrange
multiplier $\lambda$. This approximation is plausible in a spatially ordered
magnet, but in a disordered one the dependence of $\lambda_i$ on the
position may be important. In what follows we
introduce two Lagrange
multipliers: $\lambda_1$ for the sites connected by the defect bond, and
$\lambda$ for all other sites.

Performing the mean-field decoupling\cite{sark} in Eq.~(\ref{h}), one obtains
\begin{equation}
 H = H_0 + H_{int} .\label{h1}
\end{equation}
Here $H_0$ is the Hamiltonian of the undoped ferromagnet,
\begin{equation}
H_0 =-2Q_fJ\sum_{<ij>}(B_{ij} + B_{ij}^\dagger ) +
\lambda\sum_i(a_i^\dagger a_i + b_i^\dagger b_i ), \label{h0}
\end{equation}
where the mean-field amplitude $Q_f = <B_{ij}^\dagger> = <B_{ij}>$
describes the short-ranged ferromagnetic correlations.\cite{sark,aur}
The amplitude $Q_f$, like $\lambda$, is supposed to be positionally
independet in the ordered crystal.\cite{ar}

The term $H_{int}$ in Eq.~(\ref{h1}) gives the interaction of the
spin waves with the defect bonds, $<lm>$,
\begin{equation}
H_{int} = 2W_f\sum_{<lm>}(B_{lm} + B_{lm}^\dagger)
 + (\lambda_1 - \lambda)\sum_{<lm>}(a_l^\dagger a_l
+ a_m^\dagger a_m + b_l^\dagger b_l + b_m^\dagger b_m),\label{hi}
\end{equation}
where \begin{equation}
W_f = Q_fJ + Q_1J^{\prime}.\label{w}\end{equation}
 The mean-field amplitude $Q_1 = <B_{\vec r_l,\vec r_l + \vec a}>$
describes the ferromagnetic correlations of the spins at the ends of
the frustrated bond $\vec a$. It differs from the amplitude $Q_f$ for
the unfrustrated bonds, and this difference is crucial for the properties
of the doped magnets.

Transforming Eq.~(\ref{h0}) to the momentum space, we find
\begin{equation}
H_0 = \sum_{\vec q}\omega_{\vec q}(a_{\vec q}^\dagger a_{\vec q} +
b_{\vec q}^\dagger b_{\vec q}).
\end{equation}
 The excitation
spectrum $\omega_{\vec q}$ has the form
\begin{equation}
 \omega_{\vec q} = \Delta +
JQ_f z(1 -\gamma_{\vec q}) , \label{om}
\end{equation}
where $\gamma_{\vec q} =
z^{-1}\sum_{\vec \delta} \exp(i\vec q \cdot  \vec \delta)= (\cos q_x +
\cos q_y)/2$, $z = 4$ is
the number of nearest neighbors, and $\vec \delta$ is summed over
nn vectors.
The spectrum is characterized by two temperature dependent parameters,
namely,
 the mean-field amplitude $Q_f$
 and the gap $\Delta = \lambda - JzQ_f$.
 They are  governed by the
equations\cite{ar}
\begin{equation} {1\over N}\sum_{\vec q }n(\omega_{\vec q}) = S
 \label{const1}
\end{equation}
\begin{equation}
Q_f = S - {1\over N}\sum_{\vec q} n(\omega_{\vec q})(1 - \gamma_{\vec q}),
\label{q0}\end{equation}
 which follow from the constraint (\ref{const}) and the above
definition of $Q_f$.
Here $n(\omega_{\vec q})= [exp(\omega_{\vec q}/T)- 1]^{-1}$ is the Bose
distribution function, $N$ is the
total number of spins.

 It is seen from the last equation that
$Q_f$ is equal to $S$ up to corrections
of order $(T/4JS)^2$. Then Eq.~(\ref{const1}) yields that
the gap $\Delta$ is finite and
exponentially small at low temperatures,
\begin{equation}
 \Delta  \propto T \exp(-4\pi JS^2 /T) \label{del}.
\end{equation}
The formula (\ref{cor}) for the correlation length $\xi(T)$ follows immediately
from this expression and Eq. (\ref{om}).\cite{ar}

The $2\times2$ interaction matrix for each defect bond in
Eq. (2.7) can be easily diagonalized. The
interaction Hamiltonian (2.7) is then rewritten as
\begin{eqnarray}
H_{int} =&&(2JQ_f\mu_f-W_f)\sum_{<lm>}[(a_l^\dagger
 - a_m^\dagger)(a_l-
a_m) + (b_l^\dagger - b_m^\dagger)(b_l - b_m)] \nonumber\\&& +
2JQ_f\mu_f\sum_{<lm>}[(a_l^\dagger+
a_m^\dagger)(a_l+a_m) + (b_l^\dagger + b_m^\dagger)(b_l + b_m)],
\end{eqnarray}
where $\mu_f = (\lambda_1 - \lambda + W_f)/4JQ_f$.
Performing the
Fourier-transformation of this Hamiltonian we find
\begin{equation}
 H_{int} = \sum_{\nu=1}^2 v_\nu\sum_l {1\over N}\sum_{\vec q_1 \vec q_2}
e^{i(\vec q_1 -
\vec q_2)\cdot\vec r_l}f_\nu(\vec q_1)f_\nu^\ast(\vec q_2)
 (a_{\vec q_1}^\dagger a_{\vec q_2} +
b_{\vec q_1}^\dagger b_{\vec q_2}). \label{hii}
\end{equation}
Here
\begin{equation}
 f_1(\vec q)= 1- e^{-i\vec q\cdot\vec a},~~~
f_2(\vec q) = 1 + e^{-i\vec q\cdot\vec a},\label{f}
\end{equation}
$v_1 = 2JQ_f\mu_f-W_f$, $v_2=2JQ_f\mu_f$, and
 $l$ is summed over the defect bonds.

\subsection{The single defect problem}

In this subsection we use the Green's function method to solve
the problem of spin wave scattering on a single defect bond.
Let us define the retarded Green's function by the relation
\begin{equation}
G(\vec q,\vec q_ 1, t-t_1) = -i\Theta (t-t_1)<[a_{\vec q}(t),a_{\vec
q_1}^\dagger (t_1)]>_T
\end{equation}
Here $\Theta(t)$ is the step-function and $<\cdots >_T$ denotes the
thermodynamic
average.

In a ferromagnet with one defect bond, the Fourier-transform
$G(\vec q,\vec q_1,\omega)$ is linked to the $T$-matrix by the relation
\begin{equation}
G(\vec q,\vec q_1,\omega) = G^0(\vec q,\omega)\delta_{\vec q, \vec q_1}
 + G^0(\vec q,\omega)T(\vec q, \vec q_1,\omega )G^0(\vec q_1,\omega),
\label{green}
\end{equation}
 where $G^0 (\vec q,\omega)$ is the Green's function for the pure ferromagnet,
\begin{equation}
  G^0 (\vec q,\omega) = (\omega - \omega_{\vec q})^{-1}, \label{g0}
\end{equation}
and the $T$-matrix satisfies the equation
\begin{equation}
 T(\vec q,\vec q_1,\omega)
 = V(\vec q,\vec q_1) + {1 \over N}\sum_{\vec q_2}V (\vec q,\vec q_2)G^0
(\vec q_2,\omega)
T(\vec q_2,\vec q_1, \omega),  \label{tm}
\end{equation}
 with
the interaction energy
\begin{equation}
V(\vec q, \vec q_1) =
 \sum_{\nu=1}^2 v_\nu f_\nu
(\vec q)f^\ast_\nu(\vec q_1) .
\end{equation}
The Green's function for the $b$-operators obeys the same equations
 (\ref{green}) and (\ref{tm}).

The constraint (\ref{const}) for the spins, at the ends of
the frustrated bond, can be written in
the form
\begin{equation}
 <a_l^{\dagger}a_l>_T  =
 -{1\over \pi}\int_{-\infty}^{+\infty}n(\omega){1\over N}
\sum _{\vec q_1 \vec q_2}Im G(\vec q_1, \vec q_2,\omega)
 d\,\omega = S.
\end{equation}
We put here for sake of simplicity $\vec r_l=0$.
Substituting into this equation the Green's function from Eq.
(\ref{green}),
 and taking into account that
\begin{equation}
-{1\over \pi}\int_{-\infty}^{+\infty}n(\omega)Im {1\over N}\sum_{\vec
q}G^0(\vec q,\omega) =
 {1\over N}\sum _{\vec q}n(\omega_{\vec q})
= S,
\end{equation}
 we find
\begin{equation}
 \int_{-\infty}^{+\infty}d\omega\, n(\omega) \sum_{\vec q \vec q_1}
Im G^0(\vec q,\omega)T(\vec q, \vec q_1,\omega )G^0(\vec q_1,\omega)
 = 0. \label{eqmu}
\end{equation}
In the same way, from the definition  of $Q_1$ we get the second equation
\begin{eqnarray}
 Q_1 &=& <a_l^\dagger a_m> =
Q_f\nonumber\\&& + {1\over \pi}\int_{-\infty}^{+\infty}d\omega\, n(\omega)
{1\over N^2} \sum_{\vec q \vec q_1}Im G^0(\vec q,\omega)
T(\vec q, \vec q_1;\omega )G^0(\vec q_1,\omega)
(1- e^{i\vec q \cdot \vec a}).\label{eqq}
\end{eqnarray}
When deriving this equation the constraint (\ref{eqmu}) has been taken
into account.

We now solve the set of Eqs.~(\ref{tm}), (\ref{eqmu}) and (\ref{eqq}).
The perturbation energy $V(\vec q,\vec q_1)$ is the sum of separable
terms, and therefore the integral equation~(\ref{tm}) is solved easily. The
result is
\begin{equation}
T(\vec q, \vec q_1) = \sum_\nu{v_\nu f_\nu(\vec q)f_\nu^\ast(\vec q_1)\over
D_\nu(\omega)}.\label{tm1}
\end{equation}
Here
\begin{equation}
D_\nu(\omega)= 1 - v_\nu\phi_\nu(\omega),\label{dnu}
\end{equation}
where
\begin{equation}
\phi_\nu(\omega) = {1\over N}\sum_{\vec q}
f_\nu^\ast(\vec q)
G^0 (\vec q,\omega)f_\nu(\vec q). \label{phi}
\end{equation}
The functions $\phi_\nu(\omega)$ can be rewritten with the aid of
Eqs.~(\ref{f}), (\ref{g0})
in the form
\begin{eqnarray}\label{phi1}\phi_1(\omega) &=& {1 \over 2JS}[-1 +\omega
\varphi(\omega)],\nonumber\\ & & \\ \phi_2(\omega) &=&
{1\over 2JS} + (4-{\omega\over2JS})\varphi(\omega).\nonumber
\end{eqnarray}
Here
\begin{equation}
  \varphi(\omega) = {1\over N}\sum_{\vec q}(\omega -
\omega_{\vec q})^{-1} \label{vphi}
\end{equation}
is the local Green's function for the undoped ferromagnet.
We find from (\ref{vphi}) that at small $\omega$ ($\omega \ll 4JS$)
\begin{equation}
 Re \varphi(\omega) = -{1\over 4\pi JS}\ln {4JS
 \over\mid \omega - \Delta\mid},~~
Im \varphi (\omega) =-{1\over 4JS} \Theta(\omega - \Delta)\label{vphi1}.
\end{equation}
 Eqs.~(\ref{dnu})~-~(\ref{vphi})) yield
\begin{equation}
 D_1(\omega) = \zeta_f(T) + \mu_f+
[1 -\zeta_f(T)-\mu_f]\omega\varphi (\omega),
 \label{d1}
\end{equation}
where
\begin{equation}
 \zeta_f(T) = 1 - {W_f(T)\over 2JQ} \label{zeta}.
\end{equation}
We show next that $\mu_f(T)$ is always small at small
temperatures $T\ll 4JS$, and the function
$\zeta_f (T)$ is small, if the frustration is sufficiently strong.
Thus, the function $D_1(\omega)$ is small too, when the defect bond is
frustrated. This means, that the $T$-matrix and, hence, the spin wave
scattering
amplitude are enhanced strongly in this case. This is the reason of the
unusually large effect of frustrated bonds on the properties of the
doped magnets.

We substitute now the function $T(\vec q_1,\vec q_2)$
from (\ref{tm1}) into  Eqs.~(\ref{eqmu}) and (\ref{eqq}).
Taking into account that
\begin{equation}
{1\over N}\sum_{\vec q}G^0(\vec q,\omega)f_n(\vec q)
= {1\over 2}\phi_n(\omega),
\end{equation}
 we rewrite  these equations   in terms of the
functions $\phi_\nu(\omega)$
\begin{equation}
v_1\int_{-\infty}^{+\infty}d\omega n(\omega)Im{\phi_1^2(\omega)\over
D_1(\omega)} =-v_2 \int_{-\infty}^{+\infty}d\omega\, n(\omega)
 Im {\phi_2^2(\omega)\over D_2(\omega)}.\label{mu1}
\end{equation}
\begin{equation}
Q_1 = S + {v_1\over 2\pi }\int_{-\infty}^{+\infty}d\omega \,n(\omega)
Im {\phi_1^2\over D_1(\omega)}.\label{q1}
\end{equation}
When writing Eq.~(\ref{q1}) we have replaced the parameter $Q_f$
for the pure crystal by $S$, since, as we shall see later, the
difference $S - Q_f \sim (T/4JS)^2$ is always smaller, than the
second term in the r.h.s. of this equation.

It follows from Eq.~(\ref{zeta}) and the definition (\ref{w}) of $W_f$ that
\begin{equation}
Q_1 = S{1-2\zeta_f\over t+1}, \label{qz}
\end{equation}
where $t = (J^\prime-J)/J$. Eqs.~(\ref{q1}) and (\ref{qz}) yield
\begin{equation}
{t+2\zeta_f\over t+1} = -{v_1\over 2\pi S}\int_{-\infty}^{+\infty}
d\omega n(\omega)Im{\phi_1^2(\omega)\over D_1(\omega)}.\label{qz1}
\end{equation}
After the expressions (\ref{phi1})~-~(\ref{vphi1}) for
 the functions $\phi_1(\omega)$
and $\phi_2(\omega)$ are inserted into Eqs. (\ref{mu1})
 and (\ref{qz1}), one finds
\begin{mathletters}\label{muqf}\begin{equation}
{2JS^2\mu_f \over T+ 8JS^2\mu_f } ={1\over 8S}K(T)\label{muqfa}
\end{equation}
\begin{equation}
{t+2\zeta_f\over t+1} ={1\over S}K(T),\label{muqfb}
\end{equation}
\end{mathletters}
where \begin{equation} K(T)= {1\over \pi }\int_0^\infty dxxn(4JSx)
[(\zeta + \mu_f - {x\over\pi}\ln{1\over x})^2 + x^2]^{-1}.
\label{KT} \end{equation}
When deriving these equestions we neglected terms of higher order
in the small quantities $\zeta_f$ and $\mu_f$.

In the quasiclassical limit, $S\gg 1$, we find that with
logarithmic accuracy
\begin{equation}K(T) ={T  l(T)\over8JS\mid \zeta_f(T) + \mu_f(T)\mid},
\label{KT1} \end{equation}
where \begin{equation}l(T) =1+ {2\over\pi}\arctan[{1\over\pi}\ln{1\over\zeta_f
 + \mu_f}],~
1 < l(T) \leq 2, \label{lT} \end{equation} if
 $\zeta_f + \mu_f > 0$, and \begin{equation}
  l(T) = 1 - {2\over \pi}\arctan[{1\over \pi}\ln{1\over\mid
\zeta_f + \mu_f\mid}], \end {equation}
if $ \zeta_f + \mu_f <0$.

{}From Eqs.~(\ref{muqf}) - (\ref{KT1}) we finally obtain
\begin{mathletters}\label{mq1}
\begin{equation}
\mu_f\mid\zeta_f + \mu_f\mid = { l(T)\over2}
({T\over 8JS^2})^2
(1+{8JS^2\mu_f\over T}),
\end{equation}
\begin{equation}
{t+2\zeta_f\over t+1}\mid\zeta_f + \mu_f\mid =
{Tl(T)\over 8JS^2}.
\end{equation}\end{mathletters}
Eqs.~(\ref{mq1}) have a solution with $\zeta_f + \mu_f > 0$
at both $t > 0$ and $t < 0$:
\begin{mathletters}\label{zms}\begin{equation}
\zeta_f(T)= {1\over4}\{-t[1+{T\over2JS^2(t+2)}]+
[t^2 + {Tl(T)(t+1)\over JS^2}]^{1/2}\},\label{zmsa}
\end{equation}
\begin{equation}
\mu_f = {T\over8JS^2}{t+2\zeta_f\over t+2},\label{zmsb}
\end{equation}\end{mathletters}
and
\begin{equation} D_1(0) = \zeta_f + \mu_f= {1\over4}\{-t +
[ t^2 + {Tl(T)\over JS^2}(t+1)]^{1/2}\}.\label{d2}\end{equation}
If $J^\prime<J$ ($t<0$) Eqs.~(\ref{qz}) and (\ref{zms}) yield\hfil\break
1)$T/JS^2\gg t^2$
\begin{mathletters}\label{zq1}\begin{equation}
\zeta_f = ({Tl(T)\over16JS^2})^{1/2} +
O[t^2({JS^2\over T})^{1/2}],~~ \mu_f={T\zeta_f
\over4JS^2}\ll\zeta_f\label{zq1a}\end{equation}
\begin{equation}
Q_1 = S[1-({Tl(T)\over4JS^2})^{1/2}].\label{zq1b}
\end{equation}\end{mathletters}
2) $T/JS^2\ll t^2\ll1$
\begin{mathletters}\label{zq2}\begin{equation}
\zeta_f = {\mid t\mid\over2} + {Tl(T)\over8JS^2\mid t\mid} +
 O({T^2\over J^2S^4\mid t\mid^3}), ~~
\mu_f = {T^2l(T)\over(8JS^2)^2\mid t\mid}\ll\zeta_f\label{zq2a}
\end{equation}
\begin{equation}
Q_1 = S(1 - {Tl(T)\over4JS^2\mid t\mid}).\label{zq2b}
\end{equation}\end{mathletters}
The function $Q_1$ decreases with the increase of $T$
much faster than $Q$. When $T\rightarrow 0$,
$D_1(0)=\zeta_f=\mid t\mid/2$, and $Q_1$,
like $Q$, tends to $S$. Thus, the correlations of the
spins, connected by the frustrated bond, are purely
ferromagnetic, when the frustrated exchange $J^\prime$
is lower than the threshold $J_c=J$.

The properties of the defect change drastically,
when $J^\prime> J$ ($t>0$). If $T/JS^2\ll t^2$, we obtain
\begin{mathletters}\label{zd}\begin{equation}
\zeta_f = {T\over 8JS^2}[l(T){t+1\over t} - {t\over t+2}]+
 O({T^2\over J^2S^4 t^3}),
\label{zda}
\end{equation}
\begin{equation}
 D_1(0) = {Tl(T)(t+1)\over 8JS^2t}. \label{zdb}
\end{equation}\end{mathletters}
It follows from Eqs.~(\ref{qz}) and (\ref{zd}) that $Q_1(T\rightarrow 0)
=S/(t+1)<S$. Thus, when $J^\prime>J$, the ferromagnetic correlations
decrease with the increase of $J^\prime$. At large $J^\prime\gg J$,
the correlation amplitude $Q_1(0) \ll S$, i.e. the
spins are almost antiparallel. The correlation function
$Q_{\vec r,\vec r+\vec\delta}$ of two neighboring spins at
large distances $r$ from the frustrated bond is also less than
$S$, and approaches $S$ with the increase of $r$ as $r^{-4}$
(See Appendix A). Thus, the spins at the ends of the frustrated
bond act on the ferromagnetic background like a dipole, in
agreement with the classical picture.\cite{f1}

The negative solution of Eqs.~(\ref{mq1})
\begin{equation}
\zeta_f = D_1(0) =-t/2 +
 O(T/4JS^2),~~t\ll 1,\end{equation}
 which appears, when
$t > 0$, resembles the bound state solution, obtained in Ref.~\onlinecite{gins}
for a frustrated bond in a 3D ferromagnet. Indeed, the $T$-matrix
(\ref{tm1}) in this case
diverges at the {\it negative} energy $\omega = -2\pi JSt/\ln(1/t)$,
which is the condition of a bound state  formation.\cite{cal}
At $t\approx 1$ this state lies far from the bottom of the
spin-wave band. Hence, it almost does not alter the spin-wave spectrum.
In contrast, the dipole-type state  affects strongly
the properties of the ferromagnet\cite{ahar,vill}. Therefore
we consider in what follows only the dipole-type states.

Eqs.~(\ref{muqf}) can be solved also at arbitrary spins.
 If $\mid t\mid\ll1$,
the function $\mu_f\ll \zeta_f$, and $\zeta_f = D_1(0)$
is governed by the equation:
\begin{equation}
\zeta_f^2(2\zeta_f + t) = 0.524S ({T\over4JS^2})^2.\label{z1}$$
\end{equation}
This equation  yields \hfil\break
1)$ J^\prime < J, ~~\mid t\mid \gg (T/4JS)^{2/3}$
\begin{equation}
\zeta_f(T) = {\mid t \mid\over 2} + 1.05
{1\over St^2}
({T\over 4JS})^2\label{z2} + O[({T\over4JS^2})^4{1\over \mid t\mid^5}],
 \end{equation}
2) $ \mid t \mid \ll (T/4JS)^{2/3}$
\begin{equation}
\zeta_f(T) = {0.645\over S^{1/3}}({T\over 4JS})^{2/3} + O(\mid t\mid),
 \label{z3}
\end{equation}
3) $J^\prime > J , ~~ 1\gg t \gg (T/4JS)^{2/3}$
\begin{equation}  \zeta_f(T) ={0.724 S^{1/2}\over t^{1/2}}
{T\over 4JS^2} +O[({T\over 4JS^2t})^2]. \label{z4} \end{equation}
Finally, if $t\geq 1$, the function $D_1$ is given by Eq. (\ref{zdb})
with $l(T)=2$, up to small terms of the order of $\ln^{-1}4JS/T$.
Thus, at $t>0$ the main features of the functions $\zeta_f(T)$
and $Q_1(T)$ are the same, as for large spins. In other words,
the dipole-like picture of the frustrated bond defect is
valid not only in the classical limit, but rather at
arbitrary spins.

\subsection{The renormalization of the spin-wave spectrum}

Averaging Eq.~(\ref{green}) over the distribution of the defects and the
orientations of the defect bonds, we find the renormalized spin wave
spectrum $\epsilon_{\vec q}$. To first order in the defect concentration
 $x$ the configurationally averaged Green's function is given as
\begin{equation}
 G^{-1}(\vec q, \omega) = \omega - \omega_{\vec q} - \Sigma (\vec q, \omega),
\end{equation}
where the self-energy is given by $\Sigma (\vec q,\omega) = x T(\vec q,\vec q
;\omega)$.
Thus, \begin{equation}
 \epsilon_{\vec q} = \omega_{\vec q} + xT(\vec q,\vec q
;\omega_{\vec q}).\label{eps}
\end{equation}
In the long-wave limit ($\epsilon _{\vec q}
\ll T/ \ln(4JS/\epsilon_{\vec q})$), we obtain for $\epsilon_{\vec q}$
\begin{equation}
\epsilon_{\vec q} = \Omega_{\vec q}
[1 -{x\over D_1(0)}]  + \Delta_1(T), \label{spect}
\end{equation}
where $\Omega_{\vec q} = 4JS(1 -\gamma_{\vec q})$.
The properties of the renormalized  gap $\Delta_1$ will be discussed later.

When the frustration is weak, we obtain $D_1(0) \approx \mid t\mid
\approx 1$. The renormalization of the spin-wave stiffness is of the
order of $x$ and temperature independent. At strong frustrations one has
$D_1(0)\approx \zeta_f (T) \ll 1$. The
renormalization is enhanced, and the enhancement increases with the
decrease of the temperature (see Eqs. (\ref{zq1}) - (\ref{zd}),
(\ref{z2}) - (\ref{z4})).

It follows from Eqs. (\ref{spect}) and (\ref{zdb}) that for
strongly frustrated bonds ($J^\prime > J$) the renormalization of the stiffness
$\rho_s$
can be written as
\begin{equation}
{\rho_s (0) - \rho_s (x) \over \rho_s(0)}=  {U_f\over T}x,
\label{ros}\end{equation}
where $U_f = 8JS^2 t/l(T)(t+1)$.
Even  a small defect concentration  $x < {T /U_f } \ll 1$
reduces significantly the stiffness.

\subsection{ The correlation length}\label{sec:ferD}

The gap in the spin wave spectrum of a pure
 ferromagnet is governed, as mentioned above, by the constraint
 (\ref{const}).
 A self-consistent way to obtain the gap
 in a doped  {\it configurationally averaged} crystal is to impose on
the spectrum (\ref{spect}) the constraint (\ref{const}), averaged
over the defect
distribution. Since the defects under consideration preserve the value of
the spins, we obtain the same Eqs.~(\ref{const1}), (\ref{q0})
 as for the undoped crystal,
with $\omega_{\vec q}$ replaced by $\epsilon_{\vec q}$.
The gap $\Delta_1$ is then given by Eq ~(\ref{del})
with the exchange energy $J$ replaced by the renormalized
value $J(1 - x/D_1(0))$. Thus, the  correlation length in the doped
crystal is given by
\begin{equation} \xi (T,x) \propto \exp [{2\pi JS^2\over T}(1
 - {x\over D_1(0)})].\label{cor1}\end{equation}
At a given $x$ the ratio $\xi (T,x)/\xi (T,0)$ decreases rapidly
with the decrease
of the temperature.
 Even though the expression (\ref{spect}) is valid only at small concentrations
 $x \ll D_1(0)$, the renormalization of the correlation length may be
 exponentially large, if $x\gg D_1(0) (T/2\pi JS^2)$.

\section{Defect bonds in antiferromagnets}\label{sec:afer}
\subsection{The Hamiltonian}

We shall consider a bipartite antiferromagnet on a square lattice. Only
nearest neighbor interactions are included in the Hamiltonian, and it is
supposed, as before, that there is a small number of defect bonds with exchange
energy
$J^{\prime}\not = J$, where $J$ is the host exchange energy. The
Hamiltonian is given by
\begin{equation}
H = \sum_{<ij>}J_{ij}\vec S_i\cdot\vec S_j,
\end{equation}
where the couplings $J_{ij}$ are equal to $J$ or $-J^\prime$.
The transformation to Schwinger bosons for the spins on the sublattices
$A$ and $B$ is as follows:
 \begin{mathletters}\label{sb}\begin{equation}
S_A^+ =
a^{\dagger}b, ~~ S_A^- = ab^{\dagger}, ~~ S_A^z = {1\over2}(a^{\dagger}a -
b^{\dagger}b)\end{equation}\begin{equation}
S_B^+ = -ab^{\dagger}, ~~ S_B^- =-a^{\dagger}b, ~~
S_B^z = {1\over 2}(b^{\dagger}b - a^{\dagger}a) \end{equation}
\end{mathletters}
with the constraint (\ref{const}). After the mean-field decoupling\cite{sark}
is performed,
the pure antiferromagnet Hamiltonian $H_0$ and the interaction
Hamiltonian $H_{int}$ transform to
\begin{eqnarray}
H_0 = -{JQ\over 2}\sum_{<ij>}(a_i a_j +
 a_i^{\dagger}a_j^\dagger
&+& b_i b_j + b_i^{\dagger}b_j^\dagger)\nonumber\\&+& \lambda\sum_i
(a_i^{\dagger}a_i + b_i^{\dagger}b_i),
\end{eqnarray}
\begin{eqnarray}
&&H_{int} = (W - 2JQ\mu)\sum_{<lm>}[(a_l -a_m^{\dagger})(a_m - a_l^{\dagger})
 + (b_l - b_m^{\dagger})(b_m -
b_l^{\dagger})]\nonumber\\ &&+
2JQ\mu\sum_{<lm>}[(a_l^{\dagger}+a_m)(a_m^{\dagger}+a_l) + (b_l^{\dagger}+b_m)
(b_m^{\dagger}+b_l)].\label{hin}
\end{eqnarray}
Here $W = JQ + J^\prime Q^\prime$, $\mu = (1/4JQ)(\lambda_1 - \lambda
+ W)$,
\begin{equation}
 Q= {1\over2}<a_i a_j + b_ib_j > = {1\over2}
<a_i^{\dagger}a_j^{\dagger} +
b_i^{\dagger}b_j^{\dagger}>, \label{eqq1}
\end{equation}
 and  $Q^{\prime}$ is given by
Eq.~(\ref{eqq1}) with the bonds $<ij>$ replaced by the defect bond $<lm>$.
The sums in (\ref{hin}) are over all defect bonds $<lm>$.

In momentum space the Hamiltonian $H_0$ can be diagonalized by the standard
Bogoliubov transformation \begin{mathletters}\label{btr}\begin{equation}
a_{\vec q} = u_{\vec q}\alpha_{\vec q}
+ v_{\vec q}\alpha_{-\vec q}^{\dagger}~,\end{equation}\begin{equation}
 b_{\vec q}   =u_{\vec q}\beta_
{\vec q}
+ v_{\vec q}\beta_{-\vec q}^{\dagger}~,\end{equation}
\end{mathletters}
where \begin{equation}
u_{\vec q}^2 = {JQz + \lambda
 +\omega_{\vec q}\over 2\omega_{\vec q}}, ~~v_{\vec q}^2=
 {JQz + \lambda -\omega_{\vec q}\over 2\omega_{\vec q}}.\label{uv}
\end{equation}
The quasiparticle energy has the form\begin{equation}
 \omega_{\vec q}^2
= (JzQ)^2(1-\gamma_{\vec q}^2) + \Delta^2,\label{ener}
\end{equation} where the gap is given by
 $\Delta^2 = (\lambda + JQz)^2 -(JQz)^2$.
{}From  equations (\ref{const}) and (\ref{eqq1})-(\ref{ener})
 the self-consistent
equations, obeyed by the mean-field amplitude $Q$ and the gap
$\Delta$, can be obtained: \cite{ar}\begin{eqnarray}
(JQz +
\lambda){1\over N}\sum _{\vec q}\omega_{\vec q}^{-1}(n_{\vec q}
+{1\over2})&=& S+{1\over2}~,\nonumber\\ & & \\ {Jz\over N}
\sum _{\vec q}\gamma_{\vec q}^2\omega_{\vec q}^{-1}(n_{\vec q} +
{1\over 2})& =&1.\nonumber
\end{eqnarray}
 In the limit of large spin $S\gg1$, one has the amplitude
 $Q=S(1+0.158/2S)$,
 the gap  \begin{equation}
\Delta \propto T\exp (-2\pi JQS^\ast/ T),\label{Delta}
\end{equation} and the correlation
 length \begin{equation}
\xi (T) \propto \exp (2\pi JQS^\ast/T),\label{Cor}
\end{equation} where
$S^\ast = S(1- 0.197/S)$.\cite{man1}

The interaction Hamiltonian (\ref{hin}) can be written in the momentum
representation in a simple form, \begin{equation}
 H_{int} = {1\over N}\sum_{\nu=1}^2V_\nu\sum_{\vec q \vec q_1}
<A_{\vec q}\mid\eta_\nu(\vec q)><\eta_\nu(\vec q_1)\mid A_{\vec q_1}>.
\label{Hin} \end{equation}
Here we introduced the two-component column vectors $\mid A_{\vec q}>$,
$\mid\eta_\nu(\vec q)>$, and row vectors $<A_{\vec q}\mid$,
$<\eta_\nu(\vec q)\mid$ as:
\begin{equation}
\mid A_{\vec q}> = \left(\begin{array}{c}a_{\vec q}\\ a^\dagger_{-\vec q}
\end{array}\right),~ <A_{\vec q}\mid =\left(a_{\vec q}^\dagger~,
 a_{-\vec q}\right);\label{vA}
\end{equation}
\begin{equation}
\mid\eta_\nu(\vec q)> = \left(\begin{array}{c}1\\ (-1)^\nu e^{-i\vec q
\cdot\vec a}\end{array}\right), ~~
 <\eta_\nu(\vec q)\mid = \left(1,~~(-1)^\nu e^{i\vec q
\cdot\vec a}\right),\label{eta}\end{equation}
and $V_1 = 2JQ\mu - W$, $V_2 = 2JQ\mu$.

\subsection{ The one-bond problem}
Let us define the Green's function matrix by the relation
\begin{equation}
G_{\alpha\beta}(\vec q,\vec q_1;t-t_1)=-i\Theta(t-t_1)
<[A_{\alpha\vec q}(t),
A_{\beta\vec q_1}^\dagger(t_1)]>_T,\label{graf}\end{equation}
where $A_{1\vec q} = a_{\vec q}, A_{2\vec q} =  a^\dagger_{-\vec q}$.
 The unperturbed Green's function matrix is as follows:
\begin{equation}
\widehat G^0(\vec q,\omega)={1\over\omega^2-\omega^2_{\vec q}}\left(
\begin {array}{cc}4JQ+\omega & 4JQ\gamma_{\vec q}\\4JQ\gamma_{\vec q}&
4JQ-\omega\end{array}\right).\label{gr0}
\end{equation}
The one-defect problem is solved in the same way, as for the ferromagnet.
 The Green's function matrix and the $T$-matrix obey
 Eqs. (\ref{green})
and (\ref{tm}), with the functions $G$ and $T$ replaced by the
$2\times2$ matrices $\widehat G$ and $\widehat T$.
 The solution for $\widehat T(\vec q,\vec q_1;\omega)$ is
\begin{equation}
\widehat T(\vec q,\vec q_1;\omega) = \sum_{\nu=1}^2{V_\nu\mid
\eta_\nu(\vec q)><\eta_\nu(\vec q_1)\mid\over D_\nu(\omega)}.
\label{tm2}\end{equation}
Here \begin{equation}
D_\nu(\omega) = 1 - V_\nu\psi_\nu(\omega),\label{Dnu}\end{equation} and
\begin{equation}
\psi_\nu = {1\over N}\sum_{\vec q}<\eta_\nu(\vec q)\mid\widehat G^0
(\vec q;\omega)\mid\eta_\nu(\vec q)>.\label{psi}\end{equation}
Using Eqs. (\ref{eta}) and (\ref{gr0}) we find
\begin{mathletters} \label{2psi}\begin{equation}
 \psi_1(\omega)=-{1\over 2JQ}[1 -
(\omega^2-\Delta^2)g(\omega)],\end{equation} \begin{equation}
\psi_2(\omega)=
{1\over 2JQ} + (16JQ- {\omega^2- \Delta^2\over2JQ})g(\omega),
\end{equation}\end{mathletters}
where the local unperturbed Green's function $g(\omega)$ is
\begin{equation}
g(\omega) = {1\over N}\sum_{\vec q}(\omega^2 -\omega_{\vec q}^2)^{-1}.
 \label{grloc}\end{equation}
The function $g(\omega)$ is easily calculated at small frequencies
$\omega\ll 4JQ$:
\begin{mathletters}
\label{rig} \begin{equation}
Re g(\omega) =-{1\over 8\pi J^2Q^2}
[\ln{8JQ\over\mid\omega^2 -\Delta^2\mid^{1/2}} -
{1\over2\pi} + O({\omega^2\over J^2Q^2})],\end{equation}
\begin{equation} Im g(\omega) = -
 {1\over (4JQ)^2}[\Theta(\omega -\Delta)
-\Theta (-\omega-\Delta)].\end{equation}\end{mathletters}
{}From Eqs. (\ref{2psi}) and (\ref{grloc}) we find the function $D_1(\omega)$,
\begin{equation}
D_1(\omega) = \zeta(T) + \mu+
(1-\zeta-\mu)\omega^2 g(\omega),\label{D1}\end{equation}
where
\begin{equation}
\zeta(T) = 1 - {W(T)\over 2JQ}. \label{zaf}\end{equation}
We see that the spin-wave scattering on
frustrated bonds is enhanced strongly, since, as we show  below,
$\zeta (T)$ and $\mu(T)$ are small.

Let us proceed to the derivation of the equations, which govern
the function $Q^\prime$ and the local Lagrange multiplier
$\mu$. In terms of the Green's function matrix, the
 local constraint in the momentum representation can be written as
\begin{equation}
\int_{-\infty}^{+\infty}d\omega \,n(\omega)
Im \sum_{\vec q\vec q_1}<\chi_{\uparrow}\mid\widehat
G(\vec q_, \vec q_1;\omega)- \widehat G^0(\vec q,\omega)\delta_{\vec q
\vec q_1}
\mid\chi_{\uparrow}>= 0,\label{caf}\end{equation}
where $\mid\chi_{\uparrow}> $ is the spin 1/2 spinor with spin up.

Equations (\ref{green}), (\ref{tm2}) and (\ref{caf}) yield
\begin{equation}  \int_{-\infty}^{+\infty}d\omega n(\omega)
Im \sum_{\nu =1}^2V_{\nu}{L_{\uparrow \nu}(\omega)
L_{\nu\uparrow}(\omega)\over D_{\nu}(\omega)} = 0.\label{muaf}
\end{equation}
Here \begin{equation}
L_{\uparrow\nu}(\omega) = L_{\nu\uparrow}(\omega)={1\over N}
\sum_{\vec q}<\chi_{\uparrow}\mid\widehat G^0(\vec q,\omega)
\mid\eta_{\nu}(\vec q)>.\end{equation}
The equation for $Q^\prime$ can be derived
in the same way.
The first step is to transform the
initial expression
\begin{equation}
 Q^\prime = {1\over N}\sum_{\vec q \vec q_1}<a_{\vec q}^\dagger
 a_{-\vec q_1}^\dagger>
e^{i\vec q\cdot \vec a}\end{equation}
with the aid of Eq.~(\ref{tm2}), and the constraint
 (\ref{caf}). We find
\begin{eqnarray}Q^\prime &=& Q -
 {1\over \pi }\int_{-\infty}^{+\infty}
d\omega n(\omega)\nonumber\\ &\times&
Im\sum_\nu{V_\nu L_{\nu\uparrow}(\omega)
\over D_\nu(\omega)}{1\over N}
\sum _{\vec q }<\chi_{\downarrow} e^{i\vec q\cdot \vec a}
-\chi_{\uparrow}
\mid\widehat G^0(\vec q,\omega )\mid \eta_\nu(\vec q)>,
  \end{eqnarray}
where $<\chi_\downarrow\mid$ is the spinor with spin down.
Taking into account the relation\\ $<\chi_\downarrow
e^{i\vec q\cdot\vec a} - \chi_\uparrow\mid = -<\eta_1(\vec q)\mid$,
we find
\begin{equation}
Q^\prime = Q + {1\over \pi }
\int_{-\infty}^{+\infty}d\omega n(\omega)Im{V_1\psi_1(\omega)\over
D_1(\omega)}L_{1\uparrow}(\omega). \label{Qpr}\end{equation}
The functions $L_{\nu\uparrow}(\omega)$ can be written in
terms of the  functions $\psi_\nu(\omega)$ and $g(\omega)$, as
\begin{mathletters}\begin{equation}
L_{1\uparrow}(\omega) = {\psi_1(\omega)\over2} +
\omega g(\omega),\end{equation}\begin{equation}
L_{2\uparrow}(\omega) = (\omega + 8JQ)g(\omega)- {1\over2}\psi_1(
\omega).\end{equation}\end{mathletters}
Using the relation between $Q^\prime$ and $\zeta$
\begin{equation} Q^\prime = Q{1-2\zeta\over1+t}\label{Qprz}
\end{equation}
 and Eqs.
(\ref{Delta}), (\ref{2psi}), (\ref{muaf}), and (\ref{Qpr}) we finally obtain
\begin{mathletters}\label{muQ}\begin{equation}
 \mu = {T\over16JQ S^{\ast2}}(1 + {8JQS^\ast\mu\over T})I(T),
\label{muQa}\end{equation}
\begin{equation} {t+2\zeta\over t+1} = {1\over Q}I(T),\label{muQb}
\end{equation}\end{mathletters}
where \begin{equation}
I(T) = {1\over \pi}\int_0^\infty dx [x^2 \coth{2JQx\over T}-x\zeta]
[(\zeta +\mu
-{x^2\over \pi}\ln{1\over x^2})^2 + x^4]^{-1}.\label{IT}
\end{equation}
In the large spin limit the integral $I(T)$ is given by
\begin{equation}
I(T) = {Tl(T)\over8JQ(\zeta+\mu)} + {1\over2}[{\pi\over2(\zeta+
\mu)\ln(\zeta+\mu)^{-1}}]^{1/2}+  O(1) + O(\ln^{-2}{1\over\zeta}),
{}~\zeta+\mu>0.\label{IT1}\end{equation}
The second term in the r.h.s. of this equation takes into account
the effect of quantum fluctuations. The slowly varying function
$l(T)$ is given by Eq.~(\ref{lT}) with $\zeta_f, \mu_f$ replaced
by $\zeta, \mu$.

Several results follow immediately. First, we see that
 $\mu $, like $\mu_f$, tends
to $0$, when $T\rightarrow 0$. Near the
threshold, when $\mid t \mid \ll 1 $, the quantum correction to
 the mean-field amplitude $Q^\prime$ exceeds the
 correction (of the order
of $1/S$)
to the function $Q$ in the undoped samples. Indeed, if
$S^{-2/3}\ll\mid t\mid \ll 1$, and $T=0$, the mean-field amplitude becomes
\begin{equation}
Q^\prime = Q - ({\pi\over \mid t \mid \ln(1/\mid t\mid)})^{1/2}.
\end{equation}
The quantum correction increases with the decrease of $\mid t\mid$  as
$(\mid t\mid \ln \mid t\mid)^{-1/2}$, in agreement with the result obtained
in the linear spin wave theory.\cite{am} However, while in the linear
spin-wave theory the quantum corrections diverge when
 $\mid t\mid\rightarrow 0$,\cite{am,ls} we find a finite value in this
case \begin{equation}
Q^\prime = Q[1 - ({\pi\over (2Q)^2\ln 2Q})^{1/3}].\label{qst}\end{equation}
It follows from Eqs.~(\ref{muQb}) and (\ref{IT1}) that the quantum corrections,
associated with the spin-wave scattering on strongly
frustrated bonds ($t\gg S^{-2/3}$), are of the order of $S^{-2}\ln^{-1}S$,
i.e. {\it smaller} than the usual $1/S$ corrections:
\begin{equation}
\zeta = ({t+1\over t})^2{\pi\over8Q^2\ln(2Q)^2},\label{zqw}\end{equation}
\begin{equation}
Q^\prime = {Q\over1+t}[1- ({t+1\over t})^2{\pi\over(2Q)^2\ln(2Q)^2}].
\label{qst1}\end{equation}
When deriving these equations we neglected terms of order unity in
Eq.~(\ref{IT1}). The effect of these terms is to replace $2Q$
in Eqs. (\ref{zqw}), (\ref{qst1}) by $2Q+\kappa$, with $\kappa\approx 1$.
Thus $\zeta\ll1 $ for any value of the spin.
This implies that
 the quantum effects are not very important even
for real spins $S = 1/2$.

Neglecting the quantum fluctuations, we find that at $T=0$, $t>0$
one has
 $\zeta(T) = 0$, and
$Q^\prime = S/(1 + t) < S$.
Like in the ferromagnet, the function $S-Q_{\vec r,\vec r +\vec\delta}$
decays as $r^{-4}$ at large distances $r$ from the frustrated bond
(see Appendix A). Hence, the spins at the end of the frustrated bond
act as a dipole, when $J^\prime>J$.

At finite temperatures Eqs.~(\ref{muQ}) and (\ref{IT1}) yield in the large
 spin limit
that at $t\gg\zeta$ one has \begin{equation}
D_1(0) = \zeta + \mu= {Tl(T)(t+1)\over8JQ^2t}.\label{D1f}\end{equation}
The function $D_1(0)$ coincides, up to terms of the order
 of $1/S$, with Eq.~(\ref{zd})
for $D_1(0)$ in a ferromagnet. The solution of Eqs. (\ref{muQ})-(\ref{IT1})
 in the
 quasiclassical limit at small
$t$ also coincides with the corresponding solutions [Eqs.~(\ref{zq1}),
(\ref{zq2})] in
a ferromagnet, and will not be given here.

So far, we discussed only the positive solution $D_1>0$ of
 Eqs.~(\ref{muQ}).
As for the ferromagnet, these  equations also have
a negative solution  at $J^\prime >J$. However, unlike
the ferromagnet, this solution does not
correspond to a bound state, since the zero temperature T-matrix
in the antiferromagnet has no pole at negative $\omega$.

\subsection{ The spin-wave spectrum and the correlation
length}\label{sec:aferC}

To leading order in $x$, the configurationally averaged
Green's function is given
by\begin{equation}
[1- \widehat G_0(\vec q;\omega)\widehat T(\vec q,\vec q;\omega)x]
\widehat G(\vec q,\omega)= \widehat G_0(\vec q;\omega).
\end{equation}
The renormalized spin-wave spectrum, which follows from the solution of
this equation, can be written in terms of the
$T(\vec q,\vec q)$-matrix elements as \begin{equation}
\epsilon_{\vec q}^2 = \omega_{\vec q}^2 +x[4JQ(T_{11}+T_{22}) +
 \omega_{\vec q}(T_{11}-T_{22})+ 4JQ\gamma_{\vec q}(T_{12}+T_{21})].
\label{eT}\end{equation}
We have at small frequencies, $\omega_{\vec q}\ll4JQ$,\begin{equation}
\epsilon_{\vec q}^2=  \omega_{\vec q}^2[1-Re{x\over D_1(\omega_{\vec q})}].
\label{eD}\end{equation}
In the small $q$ limit, when $D_1(0)\gg  (\omega_{\vec q}^2/8\pi
J^2Q^2)\ln(4JQ/
\omega_{\vec q})$, the spin-wave spectrum is
\begin{equation}
\epsilon_{\vec q}^2 = E_{\vec q}^2 [(1 - {x\over D_1(0)}]
+ \Delta_1^2 (T),\label{Eps} \end{equation}
where $E_{\vec q}^2 = (4JQ)^2 (1 - \gamma_{\vec q}^2)$.
At higher frequencies the function $D_1(0)$ can be dropped
in Eq.~(\ref{eD}), and
we find \begin{equation}
\epsilon_{\vec q}^2 = E_{\vec q}^2 +8\pi J^2Q^2x\ln^{-1}{4JQ\over E_{\vec q}}.
\label{e2}\end{equation}
The wave-vector dependence of the last term in the r.h.s.
of this equation is subtle.
The spin-wave spectrum, hence, acquires a concentration dependent gap,
the spin-wave velocity being the same as in the undoped antiferromagnet.

Like for ferromagnets, the renolmalization in the small
 $q$ region is of the order
of $x$ for weakly
frustrated bonds, and is enhanced and temperature dependent,
when  $t < 0, \mid t\mid \ll 1$ or $t>0$. For strongly
frustrated bonds ($t>0$) we have \begin{equation}
\epsilon_{\vec q}^2 = E_{\vec q}^2 (1 - {2Ux \over T})
+ \Delta_1^2(T),\label{eaf}\end{equation}
where \begin{equation}
U = {4JQ^2t\over l(T)(t+1)}.\label{Uaf}\end{equation}
The renormalized spin wave velocity is, hence, given by
\begin{equation} c(x)= 2^{3/2}JQ(1- {Ux\over T}).\label{velb}
\end{equation}

The {\it averaged} constraint (\ref{const}) governs, like for ferromagnets, the
renormalized spin-wave gap $\Delta_1(x)$. For the correlation length at
$S \gg 1$
we find Eq. (\ref{Cor}), with the exchange coupling renormalized
according
to Eq.~(\ref{Eps}):
\begin{equation} \xi(T) = C\exp[{2\pi JQS^\ast\over T}(1-{x\over2D_1(0)})].
\label{ksi1}\end{equation}
As noted above, the SBMFA gives correctly only the exponent.
The two-loop calculation of Chakravarty {\it et al.},\cite{chak} as well as
Monte Carlo simulations (Ref.~\onlinecite{man1} and references therein) and
experimental data,\cite{keim1,grev} show that the prefactor $C$ in this
equation
does not depend on the temperature
up to small terms of the order of $T/2\pi\rho_s$.
At $J^\prime>J$, we have
\begin{equation}\xi(T) = C\exp[{2\pi JQS^\ast\over T}(1-{Ux\over T})].
\label{Cor1}\end{equation}
At sufficiently low temperatures
the ratio $\xi(x,T)/\xi (0,T)$ is small even  at
small dopant concentration $ x \ll T/U \ll 1$.

\subsection{ The  phase transition in the quasi-2D antiferromagnets}
\label{sec:aferD}

The decrease of the 2D correlation length with doping causes
the rapid reduction of the N\'eel temperature in the
quasi-2D antiferromagnets.  Starting from the relation
$^1$
\begin{equation}T_N (x) \approx J_\perp \xi^2(T_N,x),
\label{TNK}\end{equation}
 Eq.~(\ref{ksi1}) yields
\begin{equation}
T_N (x) = T_N (0) [1 - {x\over D_1(T=T_N (0))}], \label{TN}
\end{equation}
where  the N\'eel temperature for the undoped antiferromagnet
is given by
\begin{equation}T_N (0) \approx 4\pi JS^2 /\ln(J/J_\perp).
\end{equation}
Since $x$ in Eq.~(\ref{TN}) is multiplied by a large factor
$1/D_1(T_N(0))$, the N\'eel temperature is supressed rapidly,
when the dopant concentration increases.
 When $J^\prime > J$, the N\'eel temperature decreases with the
increase of doping as\begin{equation}
  T_N(x) =T_N(0)(1 - {Ux\over T_N(0)}).\label{TN1}\end{equation}
At a sufficiently large ratio $J/J_\perp$, the N\'eel temperature
extrapolates to zero at a doping concentrtion $x$
which is much smaller than the
percolation threshold.

\section{Frustrated plaquettes}\label{sec:plaq}
In this section we consider the effect of a more complicated defect,
i.e. a frustrated plaquette, on the properties of a 2D antiferromagnet. The
plaquette is formed from 4 frustrated bonds, which connect 4
neighboring spins.
To begin with, let us summarize the results obtained in
Appendix B in the harmonic spin-wave approximation. The $T$-matrix
has a pole, when the frustrated coupling, $J^\prime$, reaches the
 critical value $J_c = 0.376\,J$. Like in the case of a single frustrated
bond, the divergence of the $T$-matrix signals a local instability
of the system. When $J^\prime > J_c$, the defect gives rise to a canted
ground state.
The threshold for instability of the perfectly
aligned antiferromagnetic ground state is shifted to a weaker
value, when the extra hole frustrates four bonds in a plaquette,
rather than one bond. The reason is that in a plaquette two of
the four bonds, connecting each spin with its neighbors,
are frustrated.

There is a remarkable difference between the effect of  frustrated
plaquettes on the spin-wave  stiffness in the ferro- and
antiferromagnets. In antiferromagnets the stiffness diverges
when $J^\prime \rightarrow J_c^ -$. In ferromagnets the divergent
term in the $T$-matrix scales as $q^4$, and hence the
stiffness passes smoothly  the singular point
$J^{\prime} = J_c$. Therefore we employ the SBMFA
to consider frustrated plaquettes only in
antiferromagnets.

The interaction Hamiltonian we wish to treat is the following
\begin{equation}
H_{int} = W_{pl}\sum_{<lm>}(a_la_m + a_l^\dagger a_m^\dagger)
+ (\lambda_1 - \lambda)\sum_{l=1}^4 a_l^\dagger a_l \label{Hint}.
\end {equation}
Here the first sum $<lm>$ runs over the frustrated bonds $<12>,<13>,<24>,<34>$;
$\lambda_1$ is the Lagrange multiplyer for the spins 1-4 at the
corners of the plaquette. The energy $W_{pl} = J^\prime Q_{pl} + JQ$,
and $Q_{pl}$ is the correlation amplitude of the frustrated bonds.
Evidently, $Q_{pl}$ is the same for all 4 frustrated bonds in the
 plaquette.

The interaction energy matrix
\begin{equation} \widehat V  =  \left( \begin{array}{cccc}
\lambda_1 - \lambda & W_{pl} & W_{pl} & 0 \\ W_{pl} &
\lambda_1 - \lambda & 0 & W_{pl} \\
W_{pl} & 0&\lambda_1 - \lambda  & W_{pl} \\ 0 & W_{pl} & W_{pl}
 & \lambda_1 - \lambda
\end{array}\right) \end{equation}
is diagonalized by the unitary transformation (B3). The
 Fourier-transformed interaction Hamiltonian (\ref{Hint}) becomes
\begin{equation} H_{int} =\sum_{i=1}^4w_i\sum_{\vec q \vec q_1}
<A_{\vec q}\mid y_i(\vec q)><y_i(\vec q_1)\mid A_{\vec q}>.
\label{Hin1}\end{equation}
Here the two-component column (row) vector $\mid A_{\vec q}>$
($<A_{\vec q}\mid$) is given by Eq.~(\ref{vA}),
 $w_1 = w_3 =JQ\mu_{pl}- W_{pl}/2, ~~ w_2 = JQ\mu_{pl} - W_{pl}, ~~
w_4 = JQ\mu_{pl} = (\lambda_1 - \lambda + 2W_{pl})/4$,
\begin{equation} \mid y_i(\vec q)> =\left(\begin{array}{c}
l_i(\vec q) \\ m_i(\vec q)\end{array}\right),~~~
<y_i(\vec q)\mid = \left(l_i^\ast (\vec q), ~~~m_i^\ast(\vec q)\right),
\label{yq}\end{equation}
where \begin{mathletters}\label{lm}\begin{equation}
 l_1(\vec q)= l_3(\vec q) =1 - e^{-i(q_x +q_y)}, ~~
m_1(\vec q) = -m_3(\vec q) = e^{-iq_x} - e^{-iq_y}\label{lma}
\end{equation}
\begin{equation} l_2(\vec q) = l_4(\vec q) = 1+ e^{-i(q_x + q_y)}, ~~
m_2(\vec q) =-m_4(\vec q)= - e^{-iq_x} - e^{-iq_y}.\label{lmb}\end{equation}
\end{mathletters}
The equation for the one-defect $T$-matrix can be solved as before.
The result is \begin{equation}
\widehat T(\vec q,\vec q_1;\omega) = \sum_{i=1}^4w_i{\mid y_i(\vec q)>
<y_i(\vec q_1)\mid\over \Lambda_i(\omega)}.\label{tm3}\end{equation}
Here
\begin{equation} \Lambda_i(\omega) = 1 - w_i\Psi_i(\omega),
\label {L1}\end{equation} where
\begin{equation} \Psi_i(\omega) = {1\over N}\sum_{\vec q}
<y_i(\vec q)\mid\widehat G^0(\vec q, \omega)\mid y_i(\vec q)>.
\label{Psi}\end{equation}
It is straightforward to calculate the functions $\Psi_i(\omega)$
substituting expressions (\ref{gr0}), (\ref{yq}), (\ref{lm})
 into Eq.~(\ref{Psi}).
This yields \begin{mathletters}\label{Psi4}\begin{equation}
\Psi_1(\omega) = \Psi_3(\omega) = {16JQ\over N}\sum_{\vec q}
{1 - \cos q_x\cos q_y\over\omega^2 - \omega_{\vec q}^2},\label{Psi4a}
\end{equation} \begin{equation}
\Psi_2(\omega) = {8JQ\over N}\sum_{\vec q}{\sin^2\,q_x + \sin^2\,q_y
\over\omega^2 - \omega_{\vec q}^2},\label{Psi4b}\end{equation}
\begin{equation} \Psi_4(\omega)= {8JQ\over N}\sum_{\vec q}
{2 + \cos^2\,q_x + \cos^2\,q_y + 4\cos\,q_x\cos\,q_y \over
\omega^2 - \omega^2_{\vec q}}\label{Psi4c}.\end{equation}
\end{mathletters}
Setting $\mu_{pl} = 0$ in Eq.~(\ref{tm3}) recovers in the quasiclassical
limit the $T$- matrix expression derived in Appendix B, for the
plaquette defect in the long range ordered antiferromagnet at zero
temperature.

Converting the sums over $\vec q$ into integrals over the first
Brillouin zone in the reciprocal lattice, one finds numerically
the values \begin{equation}
\Psi_1(0) = -1.273/JQ;~~~ \Psi_2(0) =
-0.727/JQ. \label{npsi}\end{equation}
Hence,
 \begin{mathletters}\label{D12}\begin{equation}
\Lambda_1(0) =\Lambda_3(0) = 0.124 + 0.876\zeta_{pl} +1.273\mu_{pl},
\label{D12a}\end{equation}
\begin{equation} \Lambda_2(0) = \zeta_{pl} + 0.727\mu_{pl},
\label{D12b}
\end{equation}\end{mathletters}
where \begin{equation}\zeta_{pl} = 1- 0.727{W_{pl}\over JQ}
\label{zpl}\end{equation}
Next we show that $\zeta_{pl}$ and $\mu_{pl}$, like $\zeta$ and $\mu$
in the previous section,
are small at low temperatures,
if the frustration of the bonds in the defect plaquette is
sufficiently strong, i.e. if $J^\prime >J_c$, or
 $\mid J^\prime - J_c\mid/J_c\ll 1$.
Hence, $\Lambda_1(0)\gg \Lambda_2(0)$, i.e. the first and third
modes in Eq.~(\ref{tm3}) may be neglected when considering the
effect of the frustrated plaquette.

Given the expressions for the $T$-matrix and the functions $\Psi_i(\omega)$,
we can derive, as before, the equations for the
functions $\zeta_{pl}(T)$ and $\mu_{pl}(T)$. We find
\begin{equation}
\int_{-\infty}^{+\infty}d\,\omega\, n(\omega)\{w_2Im{[\Psi_2(\omega) +
4\omega g(\omega)]^2\over\Lambda_2(\omega)} + w_4Im{[\Psi_4(\omega)+
8\omega g(\omega)]^2\over \Lambda_4(\omega)}\} = 0,\label{mupl}
\end{equation}
\begin{equation}
Q_{pl}= Q + {w_2\over8\pi}\int_{-\infty}^{+\infty}d\,\omega\,n(\omega)
Im{\Psi_2(\omega)[\Psi_2(\omega) + 4\omega g(\omega)]\over
\Lambda_2(\omega)}.\label{Qpl} \end{equation}
Taking into account that at low frequencies, $\omega\ll4JS$, the functions
$\Psi_2(\omega)$ and $\Psi_4(\omega)$ can be simplified as
\begin{equation} \Psi_2(\omega) = \Psi_2(0) +
 {\omega^2\over JQ}g(\omega), ~~\Psi_4(\omega) = 64JQg(\omega),
\end{equation}
we find from Eqs.~(\ref{mupl}) and (\ref{Qpl}):
\begin{mathletters}\label{muQpl}\begin{equation}
\mu_{pl} = {T\over16JQS^{\ast2}}(1 + {8JQS^\ast\over T})I_1(T),
\label{muQpa}\end{equation} \begin{equation}
{t + \zeta_{pl}(j_c^{-1} + 1)\over t +1} =
{1\over Q}I_1(T), \label{muQpb}\end{equation}
\begin{equation} Q_{pl}(T) = Q {j_c-\zeta_{pl}(T)(1 +j_c)\over j_c(t+1)}.
\label{qj}
\end{equation} \end{mathletters}
Here  $t = (J^\prime -J_c)/J_c$,
 $j_c = J_c/J=0.376$, and
\begin{eqnarray} I_1(T)& =& {1\over\pi}\int_0^{\infty}d\,x\,x^2
\nonumber\\
&\times&
\coth{2JQx\over T}\{[\zeta_{pl} + {\mu_{pl}\over1+j_c}
-{1+j_c\over\pi}x^2\ln{1\over x^2}]^2 + (1+j_c)^2x^4\}^{-1}.
\label{I1T}\end{eqnarray}

Reasoning analogous to that given in the preceding section
shows that the quantum corrections in Eqs.~(\ref{muQpl})
are of the order of $(2S)^{-2}$, when $t>0$.
In the large spin limit the functions $\mu_{pl}(T)$
and $\zeta_{pl}(T)$ are governed by equations, similar to Eqs.
(\ref{muQ}) for one frustrated bond in an antiferromagnet,
\begin{mathletters}\label{muzpl}\begin{equation}
\mu_{pl}(\zeta_{pl} + {\mu_{pl}\over1+j_c}) = {l(T)S^\ast\over2Q}
({T\over16JS^{\ast2}})^2(1 + {16JQS^\ast\over T}),\label{a}\end{equation}
\begin{equation} {t +\zeta_{pl}(j_c^{-1} +1)\over t+1}
= {Tl(T)\over16JQ^2(1 + j_c)}.\label{b}\end{equation}\end{mathletters}
When $t<0$ and $T/4JQ^2\ll t^2\ll1$ the solution of these
equations is as follows:
\begin{equation} \zeta_{pl} = {j_c \mid t\mid\over1+j_c} +
{Tl(T)\over16JQ^2(1+j_c)\mid t\mid}, ~~\mu_{pl}\sim
{T^2\over(16JS^2)^2\mid t\mid} \ll \zeta_{pl},
\end{equation}
\begin{equation} Q_{pl} =Q(1 - {Tl(T)\over16j_c\mid t\mid JQ^2}).
\end{equation}
At positive $t$ and small temperatures $T/4JQ^2\ll t^2$, we find
\begin{mathletters}\label{zmLQ}\begin{equation}
\zeta_{pl} = {T\over16JQ^2}[{(t+1)l(T)\over t(1+j_c)} -
{Q^3t\over2S^{\ast 3}(1+t) -S^{\ast}Q^2t}],
\label{zmLQa}\end{equation}
\begin{equation}\mu_{pl} = {QT\over16JS^{\ast}}{t\over
2S^{\ast2}(1+t)-
Q^2t },\label{zmLQb}\end{equation}
\begin{equation}\Lambda_2(0) = \zeta_{pl} + {\mu_{pl}\over1+j_c}=
{t+1\over t(1+j_c)}{Tl(T)\over16JQ^2},\label{zmLQc}\end{equation}
\begin{equation}Q_{pl} = {Q\over t+1}\{1 - {T\over16j_cJQ^2}
[l(T)({1\over t} +1) -{tQ^3\over 2S^{\ast3}(1+t)
 - Q^2tS^\ast}]\}.
\label{zmLQd}\end{equation}\end{mathletters}
In the zero temperature limit, $Q_{pl} = Q$, if $t<0$,
and $Q_{pl}=Q/(t+1)<Q$, if $t>0$. Thus, at $t=0$ or $J^\prime=
J_c= 0.376J$, the ground state changes from collinear to canted.
It is argued in Appendix A that at large distances $r$ from
the defect plaquette,
the correlation amplitude $Q_{\vec r ,\vec r+\vec\delta}$ approaches
to $S$ as $r^{-6}$, i.e. the  frustrated plaquette acts on the
antiferromagnetic background at
large distances as a quadrupole.

We now average over the random distribution of the
frustrated plaquettes, and obtain the renormalized spin-wave
energy
\begin{equation}
\epsilon_{\vec q}^2 =\omega_{\vec q}^2[1
-x(1+j_c)Re{1\over \Lambda_2(\omega_{\vec q})}].
\end{equation}
In  the small $q$ limit this becomes \begin{equation}
\epsilon_{\vec q}^2 = 8J^2Q^2[1 - x{1+j_c\over\Lambda_2(0)}] +
\Delta_{pl}^2,\label{enpl}\end{equation}
where the gap $\Delta_{pl}$ should be calculated self-consistently,
as explained in Sec.~\ref{sec:ferD}.
Eqs.~(\ref{enpl}) and (\ref{zmLQc}) yield the renormalized
spin wave velocity for strongly frustrated plaquettes with $t>0$,
\begin{equation} c(x) = 2^{3/2}(1-{U_{pl}x\over T}),\label{vel}
\end{equation}
with
\begin{equation}
U_{pl}= {8JQ^2t(1 + j_c)^2\over l(T)(t+1)}.\label{upl}\end{equation}
Note that the energy $U_{pl}$ is larger than the
corresponding energy $U$ (Eq.~(\ref{Uaf})) for frustrated bonds.
 For strongly frustrated plaquettes, $t>0$, at low temperatures
$T/4JQ^2\ll t^2$, the renormalized correlation length $\xi$
and N\'eel temperature $T_N(x)$ are thus given by
\begin{mathletters}\label{cTN}\begin{equation}
\xi(T,x) = C\exp[{2\pi JS^\ast Q\over T}(1 - {U_{pl}x\over T})],
\label{cTNa}\end{equation}\begin{equation}
T_N(x) = T_N(0)[1 - {U_{pl}x\over T_N(0)}].\label{cTNb}
\end{equation}\end{mathletters}

\section{Discussion. Theory vs Experiment}\label{sec:dis}
We have generalized the SBMFA for
2D doped magnets, which allowed us to study the effect of
noninteracting arbitrary frustrated bonds on the spin-wave
spectrum and on the 2D magnetic correlation length at finite temperatures.
 We describe the defect bond
by two local parameters, namely, the mean-field amplitude and
the Lagrange multiplier. In a more rigorous treatment
local mean-field parameters should be introduced also for the
near-by bonds. However, even our simplified consideration is
sufficient to describe correctly many peculiar properties of the doped 2D
and quasi-2D antiferromagnets.

As noted in the Introduction, it is expected that
in real cuprates the effective ferromagnetic coupling $J^\prime$
generated by the localized holes is
larger than the coupling $J$,\cite{ahar}
 i.e. $J^\prime$
exceeds the spin canting threshold $J_c$ (for frustrated bonds
 $J_c=J$, while for plaquettes
$J_c=0.376J$,
in agreement with the results of the linear spin-wave theory,
see Refs.~\onlinecite{vann,am} and Appendix B).
In this case the renormalization of the antiferromagnet
spin-wave velocity
in the large spin limit is of the order of $Ex/T$, with
$E=U$ for frustrated bonds, and $E=U_{pl}$ for frustrated plaquettes
(see Eqs. (\ref{velb}), (\ref{Uaf}), (\ref{vel}) and (\ref{upl})).
Because of the large factor $E/T$ the renormalization
at low temperatures is large even at small doping level.

It follows from Eqs.~(\ref{Cor1}) and (\ref{cTNa}) that
the effect of doping upon the correlation length, like upon the spin-wave
velocity,  increases
with the decrease of the temperature. The ratio $\xi(x,T)/\xi(0,T)$
is of the order of unity if $ 2\pi\rho_sE/T^2 \ll x^{-1}$, and
is exponentially small when $( 2\pi\rho_sEx)^{1/2}\gg T\gg Ex$.
The rapid decrease of the 2D correlation length strongly reduces
the quasi-2D N\'eel temperature (see Eqs.~(\ref{TN1}) and (\ref{cTNb}).
The above results hold also for frustrated bonds in ferromagnets
(Eqs.~(\ref{ros}), (\ref{cor1}), (\ref{zdb})). In contrast,
frustrated plaquettes have only weak effects on
the properties of the ferromagnet,
the renormalization of the spin wave stiffness being of the
order of $x$ and temperature independent (Appendix B).

The effect of frustrated bonds is large
also at $J^\prime<J_c$, provided $\mid t\mid = (J_c-J^\prime)/J_c
\ll1$. However, the temperature dependence is weaker than
at $J^\prime>J_c$. The renormalization of the spin wave
spectrum in the large spin limit is proportional to
$x/T^{1/2}$ if $T\gg 4JS^2t^2$, and it tends to a
constant value $x/\mid t\mid\gg x$ when $T\rightarrow 0$
(Eqs.~(\ref{spect}), (\ref{d2}), (\ref{zq1a}), (\ref{zq2a})).

The strong effect of frustrating  defects on the spin-wave stiffness and
on the correlation length is related to the local instability leading to the
spin canting. Indeed, the spin-wave scattering amplitude at $T=\omega=0$
diverges at the local stability threshold $J^\prime = J_c$. At $J^\prime
>J_c$ the defects act as dipoles (bonds) or quadrupoles (plaquettes).
Hence, the disturbance of the background decays as a {\it power} of
 the distance from the defect rather than exponentially.
 This means that the scattering amplitude
has a zero frequency pole at any $J^\prime>J_c$.\cite{cal}
At finite temperatures the scattering amplitude is finite
but large and increases
with the decrease of the temperature. We believe that these arguments hold
for any type of frustrating defects, and therefore the above results are
valid qualitatively not only for frustrated bonds and plaquettes.

We  also calculated the quantum corrections to the local mean-field
amplitude in antiferromagnets in the large spin limit.
 The corrections are relatively
large, $\sim (S^2\ln S)^{-1/3}$, near the threshold (Eq.~(\ref{qst})),
however at stronger frustration ($J^\prime > J_c$) the corrections
are unusually small, $\sim (2S)^{-2}\ln^{-1}2S$
(Eqs.~(\ref{zqw})) and (\ref{qst1})). This result gives evidence
in favour of the classical model for strongly frustrated bonds in
2D antiferromagnets, proposed in Ref.\onlinecite{ahar}.

It follows from our theory that weakly frustrating and nonfrustrating
impurities act quite differently. The renormalization of the spin
stiffness is of the order of $x\ll 1$, and does not depend on the
temperature. Thus, the correlation length is given by the same
Eq. (1.1), as for undoped samples, with a concentration dependent
constant $A$. Similar results have been obtained
in Ref. \onlinecite{man2}
by means of the
quantum Monte-Carlo method.

Let us compare our results with the experimental findings
\cite{keim1,endo,yam,keim2,mats1,mats2}
 for the temperature and doping concentration
dependences
of the correlation length in quasi-2D antiferromagnets.
It has been shown in these papers that one should distinguish between
two kinds of impurities.
Dopants which introduce holes and,
presumably, frustrate the Cu-Cu bonds, supress
strongly the correlation length at low temperatures and reduce
rapidly the N\'eel temperature. In contrast, dopants which
introduce vacancies or excess electrons do not change the temperature
dependence of the correlation length and simply reduce the
stiffness in accord with percolation. The reduction of $T_N$ is much
more gradual than that in hole doped samples.
The results, presented
here, account
qualitatively for all these properties of doped
antiferromagnets.\cite{f2}

A remarkable difference between the magnetic properties
of the electron- and hole-doped materials was observed
in Ref.~\onlinecite{mats2}. In electron doped $Nd_{2-x}Ce_xCuO_4$
the reduction of $T_N$ and $\rho_s$ is of the same order
of magnitude. However, in the hole doped $La_2CuO_{4+x}$
the correlation length (and, hence, $\rho_s$) at
high temperatures ($T\approx 500 K\gg T_N(x)$) is almost
unchanged by doping, while $T_N$ essentially decreases.
 Our results permit
to explain this puzzle. In electron doped
samples the effect of doping does not depend on the
temperature, and the decrease of $\rho_s$, as well as of
$\xi$ and $T_N$, is of the order of $x$.
In hole doped samples the change of $T_N$ is larger than that of
$\xi(T\gg T_N)$, since the effect of doping grows
with the decrease of the temperature.

We now show that the experimental results confirm the temperature
dependence of the renormalized correlation length, derived in the
preceding sections.
 It follows from Eqs.~(\ref{Cor1}) and (\ref{cTNa}) that in antiferromagnets,
doped with strongly frustrating impurities, the quantity $T\ln (\xi/C)$
should be a linear function of $x/T$:\begin{equation}
 T\ln {\xi(T,x)\over C} = 2\pi\rho_s(1 -{E x\over T}),
\label{cg}\end{equation}
where $E$, as before, stands for $U$ or $U_{pl}$.
In Refs.~\onlinecite{keim1,endo,yam} the correlation length
was measured for three
doped samples $La_2Cu O_{4+x}$, with $T_N(x)$ = 245 K, 190 K and 90 K.
The precise hole concentrations for these samples are not given in the
references cited above. It is known, however,\cite{keim1,chen}
 that the hole
concentration varies approximately linearly with $325 - T_N(x)$.
Therefore one has an estimate for their ratios,
 $x(90):x(190):x(245) = 1:0.59:0.34$.
  The quantity $C$ can be
determined from the measurements of $\xi (T,0)$ in an undoped
sample$^2$ ( wee neglect its week temperature dependence, see
Sec.~\ref{sec:aferC}), and the stiffness constant is
$2\pi\rho_s \simeq$ 150 meV.\cite{keim1}
 Using these data, we plotted
in Fig.~1 the experimental values of $T\ln (\xi/ C)$ vs $x/T$. It is seen
that the experimental points for all three samples fall on a
single straight line, as expected from Eq.~(\ref{cg}).

According to Ref.~\onlinecite{keim1}, the linear function
 $T_N(x)$ extrapolates
to zero at $x\approx 2\%$. Thus, we can estimate the hole concentration
$x(90)\approx 1.4\%$. Then, from the slope of the straight line in
Fig.~1 we estimate the energy $E\approx 500$ meV. The theoretical
 values, which follow at $t\gg1$ from Eqs.~(\ref{Uaf}), (\ref{upl}) and
the relation \cite{keim1}
$2\pi\rho_s =1.15\,J$,
 are: $U=110$ meV, $U_{pl}$ = 410 meV. The last number is close to the
above experimental value.
Note that the extra hole in lanthanum cuprate is localized
on a region of the order of two lattice constants,\cite{chen} and hence
the real defect is more complicated than the above models. What is more, our
study is based on the assumption that the impurity holes are localized, which
is, apparently, violated at high temperatures.\cite{chen,ka} Nevertheless,
the above analysis shows that the theory of strongly frustrating defects
explains the experimental findings in samples with N\'eel order
at least semiquantitatively.

Keimer {\it et al.}\cite{keim1} also measured the temperature dependence of the
correlation length
$\xi(T,x)$  of Sr doped samples (x= 0.02; 0.03; 0.04), without N\'eel order.
Eq.~(\ref{cg}) fails to account for these experimental data even at high
temperatures, when $Ex/T < 1$. Keimer {\it et al.}\cite{keim1}
showed that in this case the correlation length
obeys the empirical relation \begin{equation}
\xi^{-1}(x,T) = \xi^{-1}(x,0) + \xi^{-1}(0,T), \label{ksik}\end{equation}
where $\xi(0,T)$ is the correlation length of the
 carrier-free sample, and $\xi(x,0)$ is the measured correlation length
at low temperatures ($\xi(x,0) = 150$, 65 and $45~\AA $
for $x= 0.02, 0.03$ and 0.04
 respectively).
They also attemped to fit Eq.~(\ref{ksik}) to the data for the above less doped
samples,
which exhibit 3D long-range order, $\xi(x,0)$ being a fitting parameter
($\xi(x,0) = 140 $
and $275~\AA$ for the $T_N = 90$ and 190~K samples respectively). The fit
is good only at relatively high temperatures. Figure~1 demonstrates
that Eq.~(\ref{cg}) describes the experimental data better
than Eq.~(\ref{ksik}),
 especially for the $T_N$=90~K sample,
although one needs only one parameter ($E$)
to fit the data for all
samples. This implies, first, that $\xi(x,0)$ for the less
doped samples is larger than the
measured $\xi(x,T)$, namely, $\xi(x,0)> 400~\AA$ for
the $T_N=90$~K sample. One cannot rule
out the possibility that $\xi(x,0)$ is infinite for the concentrations $x$
smaller than
 a certain (small) critical concentration $x_c$. This standpoint
is somewhat supported by
the numerical simulations in the zero-temperature limit.\cite{shrs,goodm}
It has been shown that the frustrated bonds destroy the 2D long-range order
 in the  Heisenberg antiferromagnet ($\xi(x,0)$ is, perhaps, finite
at any $x$ \cite{goodm}),
while in the XY model the correlations do not decay
exponentionallay, but rather as a power
of the distance.\cite{shrs} One can, thus, suggest that the small
XY-anisotropy, which exists in the
real cuprates, stabilizes the 2D, $T=0$ long-range order in doped samples at
 some small, but finite
doping concentration.

The finite value of $\xi(x,0)$ in the samples with high doping
level also affects the
renormalization of the spin-stiffness. We showed in
Sections \ref{sec:aferC} and
\ref{sec:plaq}
that the stiffness  renormalization
given by Eqs.~(\ref{eaf}) and~(\ref{enpl})
holds only in the small $q$ limit, when
\begin{equation}q^2\ll { T\over E\ln(4JS/\omega_{\vec q})},
\label{hyd}\end{equation} while at larger wave vectors the spin-stiffness is
the same as in the undoped samples. On the other hand,
the spin waves exist only within a region of size $\xi(x,T) $,
 i.e. \begin{equation}q\gg\xi^{-1}(x,T)\label{hyd1}\end{equation}
 (See Ref.~\onlinecite{man1} and
references therein). In slightly
doped samples, the inequalities (\ref{hyd}) and (\ref{hyd1}) are
always fulfilled, since the inverse
correlation length is exponentially small at low temperatures.
In strongly doped samples, a new temperature independent
length scale $\xi(x,0)$ appears. Thus, the region of well-defined
spin waves is restricted to distances smaller than $\xi(x,0)$,
while according to Eq.~(\ref{hyd}) only large wavelength
spin waves are renormalized.
At sufficiently high doping level, i.e. small $\xi(x,0)$,
the inequalities~(\ref{hyd}) and (\ref{hyd1}) are
violated, and hence the
spin-stiffness is not renormalized. This reasoning explains why the correlation
length for the $x$ = 0.02 sample is
 {\it larger} than for the $T_N$ = 90~K sample
(compare Figs.~8b and 10 in Ref.~\onlinecite{keim1}).

Substituting an estimated value of the energy $E\approx 500$ meV into
Eq.~(\ref{TN1}) for the N\'eel temperature, we find $T_N(x) = T_N(0)(1 -
bx)$, where $b\approx 20$ per $1\%$. The coefficient $b$ is
smaller than the experimental value by a factor $2.5 - 3$.
There are several reasons for this discrepancy. First, Eq. (\ref{TN1})
holds only at small $x$ when $bx\ll1$. Second, all our numerical estimates
are based on the above {\it approximate} linear relation
between the function $325-T_N(x)$ and $x$. Finally, the mean-field
equation
(\ref{TNK}) which relates the N\'eel temperature to the
2D correlation length should be improved.\cite{keim1}

As noted above, the holes in $La_2CuO_{4+x}$ are delocalized
at temperatures $T>T_N$.
The overall agreement of our theory, based on the frustration model,
 with the experimental results
for hole doped lanthanum cuprate supports the standpoint that even
delocalized holes frustrate the antiferromagnet. It would be difficult
to understand the striking difference between hole and electron
(vacancy) doping, as well as the temperature dependence of the spin
stiffness in hole doped samples, without this assumption. This conclusion
agrees with the results of our analysis\cite{ka} of the reentrant temperature
dependence
 of the sublattice magnetization in oxygen doped cuprates.
We gave evidence there
that delocalized holes frustrate the antifferomagnet, though less than
localized
holes. New efforts
are needed in order to elucidate this question.

Finally, we would like to note that, as has been shown in
Sections \ref{sec:afer} and \ref{sec:plaq}, the spin-wave velocity $c$ should
also change drastically
with hole doping. Even though we considered only 2D antiferromagnets,
we expect that the same renormalization takes place in quasi-2D
antiferromagnets at $T < T_N$ for spin waves with wave vectors in the
range $J_\perp/J \ll q^2\ll T/E\ln(4JS/\omega_{\vec q})$.
Here the first inequality guarantees that the interplane coupling does not
affect the spectrum, while the second one coincides with
Eq.~(\ref{hyd}).
Thus, we may conjecture
that  the renormalization of $c$ in hole doped quasi-2D antiferromagnets
is of the order of $JS^2/T$ in the large spin limit.
 The same conclusion follows from
a direct calculation of the spin wave scattering
on free dipoles at $T\ll T_N$.\cite{ka} Unfortunately, the
experimental information on this subject is very poor. We are aware
only of the Ref. \onlinecite{rm,rm1}, where it is shown that a small amount of
holes in the $CuO_2$ planes of   $YBa_2Cu_3O_{6+x}$ renormalizes
strongly the spin-wave
velocity. Neither the temperature nor the concentration dependences of
the spin-wave velocity has been studied.
 We hope that our results will stimulate new experiments in this field.
\acknowledgments
I acknowledge A.~Aharony for reading the manuscript and for many
helpful discussions and comments. I would also like to thank
A.~B.~Harris and A.~Auerbach for stimulating discussions.
This work was supported by the Wolfson Foundation and the
U.S.-Israel Binational Science Foundation.

\appendix
\section{The correlation function
$Q_{\lowercase{\vec r,\vec r + \vec\delta}}$}
In this appendix we calculate the correlation of the spins at
the ends of a bond, $\vec \delta$, at a large distance from the
frustrated bond, $\vec a$.
\subsection{Ferromagnet}
We start from the expression \begin{eqnarray}
Q_{\vec r, \vec r+\vec\delta}& =& <a_{\vec r}^{\dagger} a_{\vec r+\vec\delta}>
\nonumber\\ &=& -{1\over\pi}\int_{-\infty}^{+\infty}d\,\omega\,n(\omega){1\over
N^2}
\sum_{\vec q \vec q_1}Im G(\vec q,\vec q_1;\omega)e^{i(\vec q_1-
\vec q)\cdot\vec r + i\vec q_1\cdot\vec\delta},\label{AQ}
\end{eqnarray}
and subtract the constraint equations for the spins at the ends
of the bond $\vec\delta$,
\begin{eqnarray}
&&\int_{-\infty}^{+\infty}d\,\omega\,n(\omega)\sum_{\vec q \vec q_1}
e^{i(\vec q_1- \vec q)\cdot\vec r}
Im [G(\vec q,\vec q_1;\omega) - G^0(\vec q,\omega)\delta_{\vec q\vec q_1}]
 \nonumber\\ &&=
 \int_{-\infty}^{+\infty}d\,\omega\,n(\omega)\sum_{\vec q \vec q_1}
e^{i(\vec q_1-\vec q)\cdot(\vec r+\vec\delta)}Im
[G(\vec q,\vec q_1;\omega)
-G^0(\vec q,\omega)\delta_{\vec q\vec q_1}] = 0.
\end{eqnarray}
We find
\begin{eqnarray}
Q_{\vec r,\vec r+\vec\delta} =&& S -{1\over2\pi}\int_{-\infty}^{+\infty}
d\,\omega\,n(\omega){1\over N^2}Im
\sum_{\vec q\vec q_1}e^{i(\vec q_1-\vec q)\cdot
\vec r}[G(\vec q,\vec q_1;\omega)-G^0(\vec q,\omega)
\delta_{\vec q\vec q_1}] \nonumber\\ &&\times
(1-e^{-i\vec q\cdot\vec\delta})(e^{i\vec q_1\cdot\vec\delta}-1).
\label{AQ1}\end{eqnarray}
Eqs.~(\ref{AQ1}), (\ref{green}), and (\ref{tm1}) yield
\begin{equation} Q_{\vec r,\vec r+\vec\delta}= S +
{1\over2\pi}\int_{-\infty}^{+\infty}d\,\omega\,n(\omega)
Im{1\over D_1(\omega)}[\sum_{\vec q}G^0(\vec q,\omega)(\vec q\cdot
\vec a)(\vec q\cdot\vec\delta)e^{i\vec q\cdot\vec r}]^2,
\label{AQ2}\end{equation}
where we have taken into account that at large $r$ only small
wave-vectors are important, and expand the exponents in
powers of ($\vec q\cdot\vec a$), ($\vec q\cdot\vec\delta$).
The term proportional to $v_2 = \mu_f$ gives at low temperatures
a negligible
small contribution to $Q_{\vec r,\vec r+\vec\delta}$.
At small frequencies $\omega\ll JS/r^2$, the integration over $\vec q$
in Eq.~(\ref{AQ2}) is easily performed, and we find
\begin{equation}
Q_{\vec r,\vec r+\vec\delta} = S -{W_f\over(2JS)^2}
F(\vec r)Z_f(T).\label{AQ3}\end{equation}
Here \begin{equation}
F(\vec r) = {1\over \pi^2r^4}[\vec a\cdot\vec\delta -
 {2\over r^2}(\vec r\cdot\vec a)(\vec r\cdot\vec\delta)]^2,
\label{F}\end{equation}
\begin{equation} Z_f(T) = {1\over2\pi}\int_{-\infty}^{+\infty}
d\,\omega\,n(\omega)Im {1\over D_1(\omega)}.\label{Zf}\end{equation}
It follows from Eqs.~(\ref{zq2a}), (\ref{zdb}), and (\ref{Zf})
that in the zero temperature limit
\begin{eqnarray} Z_f(T) &=& 0~ \mbox{for}\;t<0,\nonumber
 \\ Z_f(T) &=& {2JS^2t\over t+1}~ \mbox{for}\;t>0.
\end{eqnarray}
Thus, we finally obtain that $Q_{\vec r, \vec r+\vec\delta}= Q$,
when $t<0$, and \begin{equation}
Q_{\vec r, \vec r+\vec\delta} = S[1 - {t\over t+1}F(\vec r)],\label{AQ5}
\end{equation} when $t>0$.
The $r$ dependence of $Q_{\vec r, \vec r+\vec\delta}$ given by Eqs.
(\ref{AQ5}) and (\ref{F}) holds in the region
 $1\ll r \ll(4JS^2/T)^{1/2}$.
\subsection{Antiferromagnet}
We first consider a frustrated bond. In this case the function
$Q_{\vec r, \vec r+\vec\delta}$ can be written as
\begin{eqnarray} Q_{\vec r, \vec r+\vec\delta} &=& Q- {1\over\pi}
\int_{-\infty}^{+\infty}d\,\omega\,n(\omega)Im{1\over N^2}\sum_{\vec q\vec q_1}
<\chi_{\downarrow}\mid[\widehat G(\vec q,\vec q_1;
\omega) - \widehat G^0(\vec q,\omega)\delta_{\vec q\vec q_1}]
\mid\chi_{\uparrow}>\nonumber\\&\times&e^{i(\vec q-\vec q_1)\cdot\vec r+
i\vec q\cdot\vec \delta}.
\label{Qaf1}\end{eqnarray}
Taking into account the constraints for the spins at the ends of the
frustrated bonds, we rewrite this equation as
\begin{eqnarray}
Q_{\vec r,  \vec r+\vec\delta} &=& Q +{W\over2\pi}
\int_{-\infty}^{+\infty}d\,\omega\,n(\omega)Im{1\over
D_1(\omega)}\nonumber\\&\times&
{1\over N^2}\sum_{\vec q\vec q_1}
[2e^{i\vec q\cdot\vec\delta}<\chi_\downarrow\mid
\widehat G^0(\vec q,\omega)\mid\eta_1(\vec q)><\eta_1(\vec q_1)
\mid\widehat G^0(\vec q_1,\omega)\mid\chi_\uparrow> \nonumber \\
&-& <\chi_\uparrow\mid\widehat G^0(\vec q,\omega)\mid\eta_1(\vec q)>
<\eta_1(\vec q_1)\mid\widehat G^0(\vec q_1,\omega)\mid\chi_\uparrow>(1 +
e^{i(\vec q-\vec q_1)\cdot\vec\delta})]e^{i(\vec q-\vec q_1)\cdot\vec r}.
\label{Qaf2}\end{eqnarray}
We expand the expression within the square brackets
in Eq.~(\ref{Qaf2}) in powers
of ($\vec q\cdot\vec a$) and ($\vec q\cdot\vec\delta$), and
transform it into a sum of three terms
\begin{eqnarray}&& -[{1\over N}\sum_{\vec q}e^{i\vec q\cdot\vec r}
G_{11}^0(\vec q,\omega)(\vec q\cdot\vec a)(\vec q\cdot\vec\delta)]^2
 + \{{2\over N}\sum_{\vec q}e^{i\vec q\cdot\vec r}[G_{11}^0(\vec q,\omega) -
G^0_{12}(\vec q,\omega)]\}^2 \nonumber \\&&+
{4i\over N^2}\sum_{\vec q}[G^0_{11}
(\vec q,\omega)- G^0_{12}(\vec q,\omega)]e^{i\vec q\cdot\vec r}
\sum_{\vec q_1}(\vec q_1\cdot\vec a) G^0_{11}(\vec q_1,\omega)
e^{-i\vec q_1\cdot\vec r}.\label{sum3}\end{eqnarray}
 The first of these terms, when integrated over $\vec q$, transforms
in the zero frequency limit into the above function (\ref{F}).
The last two terms have no analogy in the case of ferromagnets.
Their contribution to $Q_{\vec r, \vec r+\vec\delta}$ can be
important at small $r$. However, at large distances, one has
\begin{equation}\int d\,\vec q[G^0_{11}(\vec q) -G^0_{12}(\vec q)]
e^{i\vec q\cdot\vec r}\sim \int d\,\vec q {1
-\gamma_{\vec q}\over\omega^2_{\vec q}}e^{i\vec q\cdot\vec r} = 0.
\end{equation}
Thus the behavior of the function $Q_{\vec r, \vec r+\vec \delta}$
 at large distances in antiferromagnets is the same as in
 ferromagnets: $S-Q_{\vec r, \vec r+\vec\delta}\propto r^{-4}$
in the region $1 \ll r \ll ({4JS^2/T})^{1/2}$.
Analogous reasoning holds also in the case of  frustrated
plaquettes. Then, instead of the first term in Eq.~(\ref{sum3})
we find
$$ [\sum_{\vec q}e^{i\vec q\cdot\vec r}G^0_{11}(\vec q)(\vec q\cdot
\vec a)(\vec q\cdot\vec b)(\vec q\cdot\vec \delta)]^2,$$
where $\vec a$ and $\vec b$ are unit vectors along the plaquette
sides. Hence,  at large distances the function $S- Q_{\vec r, \vec r+
\vec\delta}$ decays as $r^{-6}$.

\section{The frustrated plaquette at zero temperature.
 Holstein-Primakoff representation}
In this Appendix we show that the effect of the frustrated plaquettes on
the spin-wave spectrum is strikingly different in ferro-  and
antiferromagnets. While in antiferromagnets the renormalization diverges
when the frustrated coupling $J^\prime$ tends to the local stability
threshould, $J_c$, in ferromagnets  the renormalization is finite in
this limit, and, hence, is qualitatively the same at $J^\prime<J_c$ and
$J^\prime>J_c$.
\subsection{Ferromagnet}
We suppose that the strength of the frustrated bond $J^\prime$
is smaller than the threshold value $ J_c$ for local instability
(the value of $J_c$ is  obtained below). Then all the spins
are parallel, and
the Hamiltonian, which describes the interaction of the spin waves
with the spins at the corners of the frustrated plaquette, is
\begin{equation}
H_{int} = (J+J^\prime)S\sum_{<lm>}(a_l^{\dagger} - a_m^{\dagger})(a_m - a_l),
\label{HA1}
\end{equation}
where  $<lm>$ is summed over the bonds in the plaquette.
It follows from (\ref{HA1}) that the interaction matrix is
\begin{equation}
\widehat V = (J+J^\prime)S\left(\begin{array}{cccc}-2&1&1&0\\1&-2&0&1\\
1&0&-2&1\\ 0&1&1&-2\end{array}\right)\label{VA} \end{equation}
It is easy to verify that the matrix $\widehat P=
\widehat U\widehat V\widehat U^{-1}$, where
\begin{equation}
\widehat U = {1\over2}\left(\begin{array}{cccc} 1&-1&1&-1\\
1&-1&-1&1\\1&1&-1&-1\\ 1&1&1&1 \end{array}\right),\label{UA}
\end{equation} is diagonal, and its elements  are
$P_1 = P_3 = -(J+J^\prime)S/2, P_2 = -(J+J^\prime)S, P_4=0$.
The Fourier transform of the Hamiltonian (\ref{HA1})  is given by
\begin{equation}
H_{int} = \sum_{i=1}^3 P_i{1\over N}\sum_{\vec q \vec q_1}
x_i(\vec q)x_i^\ast(\vec q_1)a_{\vec q}^{\dag} a_{\vec q_1},
\label{HA2}\end{equation}
with \begin{mathletters} \label{X} \begin{equation}
x_1(\vec q) =(1- e^{-i\vec q\cdot\vec a})(1 + e^{-i\vec q\cdot\vec a}),
\label{Xa}\end{equation}\begin{equation}
x_2(\vec q) = (1 - e^{-i\vec q\cdot\vec a})(1 - e^{-i\vec q\cdot\vec b}),
\label{Xb}\end{equation}
\begin{equation}
x_3(\vec q) = (1+e^{-i\vec q\cdot\vec a})(1 - e^{-i\vec q\cdot\vec b}).
\label{Xc}\end{equation}\end{mathletters}
The solution of the $T$-matrix equation can be obtained in the
usual way, \begin{equation}
T(\vec q,\vec q_1;\omega) = \sum_{i=1}^3 P_i{ x_i(\vec q)x_i^\ast(\vec q_1)
\over1-P_i\Phi_i (\omega)}, \label{TA}\end{equation}
where
\begin{equation} \Phi_i(\omega) = {1\over N}\sum_{\vec q}x_i(\vec q)
G^0(\vec q, \omega)x_1^\ast(\vec q). \label{Fii} \end{equation}
It follows from Eqs. (\ref{Fii}) and  (\ref{X}) that
\begin{mathletters}\label{Fi2} \begin{equation}
\Phi_1(\omega) = \Phi_3(\omega) = 4\int {d^2\,q\over(2\pi)^2}
{(1+\cos\,q_x)(1-\cos\,q_y)\over\omega -
\omega_{\vec q}},\label{Fi2a} \end{equation} \begin{equation}
\Phi_2(0) =4\int{d^2\,q\over(2\pi)^2}{(1-\cos\,q_x)(1-\cos\,q_y)
\over\omega - \omega_{\vec q}}.\label{Fi2b} \end{equation}
\end{mathletters}
Thus, the function $T(\vec q,\vec q;\omega)$ is given in the limit of smal $q$
and $\omega$ by
\begin{equation} T(\vec q,\vec q;0) = -{4P_1 q^2\over1-P_1
\Phi_1(0)}
-{ P_2q_x^2q_y^2\over1-P_2\Phi_2(0)}.
\label{T0}\end{equation}
The second term in the r.h.s. of this equation is of higher order
in small $q$, and can be neglected at small $P_2$. However, it appears that
with the increase of $P_2$ it is the denominator of this term which approaches
to zero first, and, hence, determines the threshold for the local
instability.

The integration over the first Brillouin zone of the reciprocal lattice
yields $$ \Phi_2(0) = -0.727(J+J^\prime)/JS,~~ \Phi_1(0) =-1.363/JS.$$
Hence $P_1\Phi_1(0) = 0.682\,(J+J^\prime)/J < P_2\Phi_2(0)=
0.727\,(J+J^\prime)/J$,
and the threshold value of $J^\prime$ should be found from the
equation
$$1 - P_2\Phi_2(0) = 0.$$ Thus $J_c = 0.376J$,
which exactly coincides with the value of $J_c$, found
in Sec.~\ref{sec:plaq} in the Schwinger boson representation for the plaquette
defect in an antiferromagnet.

Note that the renormalized spin-wave spectrum \begin{equation}
 \epsilon_{\vec q}=
\omega_q + xT(q,q) = \omega_{\vec q}[1- {1.572J_c(J^\prime+J)\over J(J_c-
0.806J^\prime)}x]\end{equation}
is nonsingular at $J^\prime = J_c$.
Therefore the renormalization of the spin-stiffness in the ferromagnet
with frustrated plaquettes, unlike ferromagnets with frustrated bonds,
is small, of the order
	of $x$, at both $J^\prime < J_c$ and
$J^\prime > J_c$.
\subsection{Antiferromagnet}
 The Hamiltonian, which describes the interaction of the spin waves with
the spins at the corners of the frustrated plaquette, when $J^\prime<
J_c$, is
\begin{equation} H_{int} = -(J+J^\prime)S\sum_{<lm>}(a_lb_m +
a_l^{\dagger}b_m^{\dagger} + a_l^{\dagger}a_l + b_m^{\dagger}b_m),
\end{equation}
where $a_l (b_m)$ are the Holstein-Primakoff operators for the spins in
the sublattice $A$ ($B$).
The interaction matrix, like in ferromagnets, can be diagonalized
by the transformation (\ref{UA}). Then the Fourier transform of the
Hamiltonian is given by
\begin{equation} H_{int} = \sum_{i=1}^3 P_i{1\over N}\sum_{\vec q\vec q_1}
<c_{\vec q}\mid Y_i(\vec q)><Y_i(\vec q_1)\mid c_{\vec q_1}>.
\end{equation} Here
\begin{equation}\mid c_{\vec q}> = \left( \begin{array}{c}a_{\vec q}\\
b_{-\vec q}^{\dagger}\end{array}\right),~~~<c_{\vec q}\mid =
\left( a_{\vec q}^{\dagger}, ~ b_{-\vec q}\right),\end{equation}
\begin{equation} \mid Y_i(\vec q)> =\left(\begin{array}{c}
\alpha_i(\vec q)\\ \beta_i(\vec q)\end{array}\right), ~~~
<Y_i(\vec q)\mid = \left( \alpha^\ast_i(\vec q)~~ \beta^\ast_i(\vec q)
\right),\label{Y}\end{equation}
where \begin{mathletters} \label{alb} \begin{equation}
\alpha_1(\vec q) = \alpha_3(\vec q) = 1 - e^{-i(q_x+q_y)}, ~
\beta_1(\vec q) = -\beta_3(\vec q) = e^{-iq_x} - e^{_iq_y},
 \end{equation} \begin{equation}
\alpha_2(\vec q) = 1+e^{i(q_x+ q_y)}, ~~\beta_2(\vec q)=
e^{-iq_x} + e^{-iq_y}.\end{equation} \end{mathletters}
The next calculations are straightforward, provided we
introduce the Green's function matrix by Eq.~(\ref{graf})
with the vectors $<c_{\vec q}\mid (\mid c_{\vec q}>)$ substituted for
$<A_{\vec q}\mid (\mid A_{\vec q}>)$.

The  unperturbed Green's functions are \begin{equation}
\widehat G^0(\vec q,\omega) ={1\over \omega^2 -\omega^2_{\vec q}}
\left(\begin{array}{cc}4JS + \omega &-4JS\gamma_{\vec q}\\
-4JS\gamma_{\vec q}& 4JS -\omega\end{array}\right).
\end{equation}
Like in the ferromagnet, only the mode $i=2$ is relevant at
$J^\prime$ near the threshold $J_c= 0.376J$. Therefore the
single-defect  $T$-matrix can be written as
\begin{equation} \widehat T(\vec q,\vec q_1;\omega) =
{-(J+J^\prime)\mid Y_2(\vec q)><Y_2(\vec q_1)\mid\over 1 -P_2R(\omega)},
\label{TB}\end{equation} where
\begin{equation} R(\omega) = {8JS\over N}
\sum_{\vec q}{\sin^2q_x+
\sin^2q_y\over \omega^2 - \omega^2_{\vec q}}.
\end{equation}
The function $R(\omega)$ coincides up to the quantum corrections with
the function $\Psi_2(\omega)$ from Eq.~(\ref{Psi4b}). Thus
$R(0) = 0.727(1 + J/J^\prime)$.
The Eqs. (\ref{TB}), (\ref{Y}), and (\ref{alb}) together with
the relation,\cite{ks}
\begin{eqnarray}\epsilon_{\vec q}& =& \omega_{\vec q}  +
{x\over2\omega_{\vec q}}[4JS(T_{11}(\vec q,\vec q) + T_{22}(\vec q,\vec q))
+\omega_{\vec q}(T_{11}(\vec q,\vec q) - T_{22}(\vec q,\vec q))\nonumber\\&-&
4JS\gamma_{\vec q}(T_{12}(\vec q,\vec q) + T_{21}(\vec q,\vec q))]
\end{eqnarray}
yield the renormalized spin-wave spectrum,
\begin{equation}
\epsilon_{\vec q} = \omega_{\vec q}[1 - {1.76(J+J^\prime)x\over J_c-
J^\prime}].\end{equation}
The remarkable difference between the effect of the frustrated plaquettes
on the spin-wave spectrum in the ferro- and antiferromagnets is
now evident.

\figure{ FIG.~1. $(T/2\pi\rho_s)\ln (\xi/C)$ as a function of
 $2\pi\rho_s x/x(90)T$ for three samples\hfil\break  $La_2CuO_{4+x}$ with
$T_N$=
245 K ($\times$), 190 K ($\circ)$, and 90 K ($\bullet$). The points
are obtained from the experimental
 data (Refs.~\onlinecite{keim1,endo,yam}), as discussed
in the text. The straight solid line shows the fit to Eq.~(\ref{cg}).
 The dotted and dashed lines represent fits to Eq.~(\ref{ksik}) for
 the $T_N= 90$ and 190~K samples ($\xi(x,0) = 140$ and
$275~\AA$, Ref.~\onlinecite{keim1}).\label{fig1}}


\begin{references}
\bibitem{f1} The canting angle of the spins around the dipole
decay as $r^{-1}$ (Ref.~\onlinecite{ahar,vill}). Since $Q_{\vec r,\vec r
+\vec\delta}
\sim (1/2)\mid\vec S_{\vec r} +\vec S_{\vec r+\vec\delta}\mid$
(Ref.~\onlinecite{aur}), the difference $S-Q_{\vec r,\vec r + \vec\delta}$
decays with the increase of $r$ as $r^{-4}$.
\bibitem{f2} It should be emphasized, however, that in 2D antiferromagnets
the bond defect renormalization of the spin stiffness
differs in some important respects from the site defect renormalization
In the later case, e.g., a low-frequency logarithmic divergence
in the spin-wave stiffness was identified.\cite{hk}
\end{references}
\end {document}